\documentclass[11pt,oneside,a4paper,reqno]{amsart}%
\usepackage[foot]{amsaddr}
\usepackage[version=3]{mhchem}
\usepackage{booktabs}
\usepackage[latin1]{inputenc}
\usepackage{upgreek}
\usepackage[pdftex]{graphicx}
\usepackage{amsmath}
\usepackage{amssymb}
\usepackage[sf,SF]{subfigure}

\usepackage{microtype}
\usepackage{cite}
\usepackage{hyperref}

\usepackage{float}

\renewcommand{\thefootnote}{\fnsymbol{footnote}}

\newcommand{\beginsupplement}{%
        \setcounter{table}{0}
        \renewcommand{\thetable}{S\arabic{table}}%
        \setcounter{figure}{0}
        \renewcommand{\thefigure}{S\arabic{figure}}
        \setcounter{equation}{0}
        \renewcommand{\theequation}{S\arabic{equation}}%
     }

\title[\ce{H2O} Dissociation on $\alpha$-A\MakeLowercase{l}$_{2}$O$_{3}\text{(1}\bar{\text{1}}\text{02)}$]{Characterization of Water Dissociation on $\boldsymbol\alpha$-A\MakeLowercase{l}$_{\text{2}}$O$_{\text{3}}$$\text{(1}\bar{\text{1}}\text{02)}$: Theory and Experiment$^{*}$}
\author{Jonas Wirth$^{1}$}
\address{$^{1}$Institute of Chemistry, Karl-Liebknecht Stra{\ss}e 24-25, D-14476, Potsdam, Germany}
\author{Harald Kirsch$^{2}$}
\address{$^{2}$Fritz Haber Institute of the Max Planck Society, Faradayweg 4-6, 14195 Berlin, Germany}
\author{Sebastian Wlosczyk$^{2}$, Yujin Tong$^{2}$, Peter Saalfrank$^{1}$, R.\ Kramer Campen$^{2}$}

\begin{document}

\maketitle
{\let\thefootnote\relax\footnotetext{$^{*}$ For the published version see, Wirth et al (2016) Phys Chem Chem Phys, 18, 14822 \href{http://dx.doi.org/10.1039/C6CP01397J}{(doi:10.1039/C6CP01397J)}}

\begin{abstract}
The interaction of water with $\alpha$-alumina (\textit{i.e.}\ $\alpha$-\ce{Al2O3}) surfaces is important in a variety of applications and a useful model for the interaction of water with environmentally abundant aluminosilicate phases. Despite its significance, studies of water interaction with $\alpha$-\ce{Al2O3} surfaces other than the (0001) are extremely limited. Here we characterize the interaction of water (\ce{D2O}) with a well defined $\alpha$-\ce{Al2O3}(1\=102) surface in UHV both experimentally, using temperature programmed desorption and surface-specific vibrational spectroscopy, and theoretically, using periodic-slab density functional theory calculations. This combined approach makes it possible to demonstrate that water adsorption occurs only at a single well defined surface site (the so-called 1-4 configuration) and that at this site the barrier between the molecularly and dissociatively adsorbed forms is very low: 0.06 eV. A subset of OD stretch vibrations are parallel to this dissociation coordinate, and thus would be expected to be shifted to low frequencies relative to an uncoupled harmonic oscillator. To quantify this effect we solve the vibrational Schr\"odinger equation along the dissociation coordinate and find fundamental frequencies red-shifted by more than 1,500 cm$^{\text{-1}}$. Within the context of this model, at moderate temperatures, we further find that some fraction of surface deuterons are likely delocalized: dissociatively and molecularly absorbed states are no longer distinguishable.
\end{abstract}

\section{Introduction}

$\alpha$-Alumina surfaces are omnipresent in industrial applications in heterogeneous catalysis, optics and electronics and a useful model for surfaces of more environmentally abundant aluminosilicates \cite{Sparks13}. Whether in application or the environment, several decades of work have made clear that their properties (\textit{e.g.}\ reactivity, polarity and structure) all change dramatically on contact with water \cite{Brown01,kelb07}.

Understanding the molecular mechanism of $\alpha$-alumina/liquid water interaction is challenging because relevant processes occur over more than ten orders of magnitude in time and space. In principle, studying the interaction of small numbers of water molecules with well-defined $\alpha$-alumina surfaces in ultra-high vacuum (UHV) should remove much of this complexity. However, even for the most studied $\alpha$-\ce{Al2O3}(0001) surface\cite{cous94, elam98, hass98, liu98, nels98, toof98, eng00, niu00} gaining molecular level insight into water/alumina interaction has proven surprisingly elusive. In fact, while theoretical studies had converged on a common view of the most thermodynamically stable $\alpha$-\ce{Al2O3}(0001) surface termination in the absence of water and the molecular mechanism of water/(0001) interaction, it was only in our previous work that clear experimental evidence for the theoretically predicted water dissociation pathways was offered\cite{kirsch2014}.   

Much previous work has shown that the reactivity of metal oxide surfaces in general, and $\alpha$-alumina surfaces in particular, is a function of the coordinative undersaturation and density of atoms exposed at the surface \cite{Brown01}. A variety of previous theoretical and experimental studies have shown that, in UHV, the $\alpha$-\ce{Al2O3}(0001) surface is Al terminated and that this termination results in a layer of Al atoms that are three fold coordinated and a layer of oxygens that are doubly coordinated. While much theory predicts that the (0001) is the most stable of $\alpha$-alumina's surfaces in UHV, the next most stable surface is generally found to be the stoichiometric (1$\bar{\text{1}}$02) (\textit{e.g.}\ 0.11 J/m$^2$ less stable than (0001) by the computation of Kurita et al.\cite{kuri10}). From a computational standpoint a variety of prior workers have found the most stable (1\=102) surface structure in UHV in the absence of water to be a stoichiometric oxygen termination that produces a (1$\times$1) pattern when probed via low energy electron diffraction (LEED) \cite{kuri10,Mason10,Tougerti11,Aboud11}. This surface is composed of quintuply coordinated surface aluminums and alternating rows of triply and quadruply coordinated surface oxygens. Some prior experimental studies have found a stable surface structure of the (1\=102) surface in UHV that produces a (2$\times$1) LEED pattern and argued this surface structure is the most stable \cite{gillet1992,schildbach1993}. More recent work, however, has found that when oxygen defects are properly filled during sample preparation the thermodynamically stable surface gives a (1$\times$1) LEED pattern, consistent with calculation (see Supporting Information for detailed discussion of this point) \cite{Trainor02,kelb07,ster13}.

While the structure of the $\alpha$-\ce{Al2O3}(1\=102) surface in UHV in the absence of \ce{H2O} thus seems clear, there are to our knowledge no published studies that systematically compare experimental observables of this system with submonolayer concentrations of water and appropriate calculation. Here, we address this shortcoming by following the combined experimental/theoretical approach of our previous study of $\alpha$-\ce{Al2O3}(0001)/water interaction \cite{kirsch2014}. We prepare well-defined $\alpha$-\ce{Al2O3}(1\=102) surfaces in UHV, expose them to submonolayer water coverages and characterize the resulting water/alumina interaction using temperature programmed desorption (TPD) and the interface specific, laser-based technique vibrational sum frequency (VSF) spectroscopy. In parallel we perform a series of Density Functional Theory (DFT) calculations. Collectively the results of this approach allow characterization of the energies and frequencies of all possible surface species as well as the thermodynamics and kinetics of relevant surface processes, especially those of the water dissociation reaction. This combined approach allows us to unambiguously demonstrate that on the (1\=102) surface in UHV unimolecular water dissociation proceeds through the \textit{so-called} 1--4 channel, but that the barrier between molecular and dissociative adsorption is quite small: $\approx$0.06 eV. Explicit calculation of the wavefunction and vibrational eigenvalues for this transition demonstrate that these two states are largely indistinguishable. While all experiments and calculations are conducted with \ce{D2O} instead of \ce{H2O}, the deuteron delocalization we describe should only be enhanced when protons are explicitly considered. Such surface deuteron delocalization has not, to our knowledge, been previously described in studies of water/oxide interaction and has significant implications for temperature dependent surface reactivity, physisorption of other solutes and, from a practical point of view, appropriate levels of theory for the description of water/alumina interaction.

\section{Materials and Methods}\label{s:meth}

\subsection{Experiment~~}\label{ss:exp}
\subsubsection{UHV System and Preparation of $\alpha$-\ce{Al2O3}(1\=102) Surfaces}
The experiments were performed in an Ultra High Vacuum (UHV) chamber with a base pressure of $ 2.5 \cdot 10^{-10}$ mbar. Connected to the chamber is a turbo molecular beam source for water dosing and a quadrupole mass spectrometer for determination of surface coverage. For our sample preparation, we adopted a treatment that has been reported previously \cite{Trainor02} to form a carbon free surface with a well defined (1$\times$1)-LEED pattern (see Supplementary Information and our previous study for details of sample preparation and temperature control \cite{kirsch2014}). In our hands this procedure also results in a carbon free surface (confirmed by Auger spectroscopy) with a well defined (1$\times$1)-LEED pattern (see Supplementary Information for Auger and LEED results).

Given a well-defined $\alpha$-\ce{Al2O3}(1\=102) surface, we prepared our sample by setting the sample temperature to 400 K and continually dosing as the sample was cooled to 135 K with a ramp of 20 K/min. As we show below, this procedure produces a mixed first layer, containing both dissociatively and molecularly adsorbed water, overlain by a layer of ice. The spectral response of the ice layer is used to align our set up before each measurement. Note that this procedure differs significantly from that employed in preparing the $\alpha$-\ce{Al2O3}(0001).  There, as discussed previously \cite{kirsch2014}, it was necessary to dose  continually from 450 K to 300 K by using a ramp of 10 K/min, followed by staying for 20 minutes at 300 K in order to generate significant coverages of dissociatively adsorbed water. Then the preparation cycle was finished by continually dosing while cooling from 300 K to 140 K with a ramp of 10 K/min. These qualitative differences clearly suggest the (1\=102) surface is more reactive than the (0001) with respect to water dissociation. This assessment of relative reactivities is consistent with prior work \cite{kelb07}.

Note that all VSF spectra described below were collected at sample temperatures of 135 K. The experimental temperature dependent stability we describe was probed by rapidly annealing a sample from 135 K to the indicated temperature (using a heating ramp of 100 K/min) and, after reaching the target temperature cooling it at the same rate back to 135 K for analysis.  

\subsubsection{VSF Spectrometer}
In a VSF measurement pulsed infrared and visible lasers are overlapped spatially and temporally at an interface and the output at the sum of the frequencies of the incident fields is monitored. The resulting sum frequency emission is interface specific by its symmetry selection rules and is a spectroscopy because emission increases dramatically if the frequency of the incident infrared is tuned to a vibration of a moiety at the interface \cite{kirsch2014,zhu87}. In our VSF spectrometer 50 fs, 30 - 40 $\mu$J, infrared pulses centred at 2700 cm$^{\text{-1}}$ with a bandwidth of 190 - 230 cm$^{\text{-1}}$ (FWHM) are spectrally overlapped with 25 $\mu$J pulses centred at 800 nm (12,500 cm$^{\text{-1}}$) with a bandwidth of 8 cm$^{-1}$ and the emitted sum frequency field detected using an ICCD camera. In our experimental configuration incident IR and VIS beams are coplanar and this plane is normal to the surface. All reported measurements were conducted in the \textit{ppp} configuration (the polarizations of all beams parallel to the plane of incidence) and incident VIS/IR were at 75/70 degrees with respect to the surface normal. For all experiments we used D$_{2}$O instead of H$_{2}$O because our IR source has more energy at OD stretch than OH stretch frequencies. As we discuss in detail below, the novel features of \ce{D2O} adsorption on this surface, \textit{e.g.}\ the small barrier between molecular and dissociated forms, seem likely to be only enhanced in the presence of \ce{H2O}. 

\subsubsection{Analysis of VSF Spectra}
The theoretical background of VSF spectroscopy has been given in detail in previous publications of us and others \cite{kirsch2014,kirsch2014controlling,tong2015optically,zhu87,hess00} and is therefore only summarised here (see Supplementary Information for a complete description). The intensity of the emitted sum frequency field ($I_{\text{VSF}}$) is related to the intensities of the incident IR and VIS fields,
\begin{equation}
	I_{\text{VSF}}\propto |\chi^{(2)}|^{2}I_{\text{VIS}}I_{\text{IR}}
\end{equation}
in which $\chi^{(2)}$ is the macroscopic nonlinear optical susceptibility and contains, in its dependence on the frequency of the incident IR field, the vibrational spectrum of interfacial moieties. To extract this vibrational spectrum quantitative line shape analysis is necessary. We perform such an analysis here following prior workers \cite{bain91, buss09} and take $\chi^{(2)}$ to be a coherent superposition of a nonresonant background with any material resonances. If each resonance is homogeneously broadened and dynamical effects are small (assumptions justified in detail in our prior work\cite{kirsch2014}) $\chi^{(2)}$ can be written (in which $\tilde{\nu}_{\text{ir}}$ is the infrared frequency):
\begin{eqnarray}
\chi^{(2)} & \propto & \chi^{(2)}_{\text{NR}} + \chi^{(2)}_{\text{R}} \label{e:line1}\\
 & \propto & \left|A_{\text{NR}} \right|\ e^{i\phi_{\text{NR}}} +\ \sum_{q} \frac{A_{q}}{\tilde{\nu}_{\text{ir}}-\tilde{\nu}_q + i\Gamma_q} \label{e:line2}
\end{eqnarray}
Where $A_{q}$, $\tilde{\nu}_q$ and $2\Gamma_q$ are the complex amplitude, center frequency and line width of the $q^{\text{th}}$ resonance and $\left|A_{\text{NR}} \right|$ and  $\phi_{\text{NR}}$ are the nonresonant amplitude and phase. In fitting this model to the data we followed standard procedures.

Given a measured macroscopic nonlinear susceptibility in a laboratory fixed reference frame (IJK), $\chi^{(2)}$ can be written as the sum ($N$) of molecular responses ($\beta^{(2)}$) within a molecular fixed reference frame (ijk) multiplied by the appropriate ensemble average transformation matrix\cite{wang05}.
\begin{equation}
\chi^{(2)}_{IJK} = N\sum_{i,j,k}\left\langle(\hat{I}\cdot\hat{i})(\hat{J}\cdot\hat{j})(\hat{K}\cdot\hat{k}) \right\rangle\beta^{(2)}_{ijk}
\end{equation}
Given a known molecular orientation and molecular nonlinear optical response, this formalism allows the calculation of the expected $I_{\text{VSF}}$ (see discussion and Supplementary Information for theoretical details). Further details of the VSF data analysis, including the constraints required to avoid nonphysical fits to the data, are given in the Supplementary Information.

Note that all presented VSF data have been normalized to account for the frequency dependence of the incident infrared energy.


\subsection{Theory}\label{ss:comp}

\subsubsection*{Computational Details}

Total energy calculations presented in this paper were performed within the formalism of Kohn-Sham density functional theory~\cite{Kohn65} as implemented in the Vienna Ab initio Simulation Package (VASP)~\cite{Kresse93a,Kresse93b,Kresse99}. A plane-wave basis was adopted using the Projector-Augmented plane Wave (PAW) scheme~\cite{Blochl94,Kresse99} and a kinetic energy cutoff of 400\,eV. Electron exchange and correlation were described using the PBE functional~\cite{Perdew96b}; in case of the water dissociation reaction (see below) also the PBE-based, range-separated hybrid functional HSE06~\cite{Paier06} was used in order to correct for the well-known underestimation of reaction barriers~\cite{Zhao05,Johnson94} with standard DFT functionals. In order to account for dispersion interaction, Grimme's semi-empirical D3 correction with a damping function according to Becke and Johnson~\cite{Grimme10,Grimme11} was added to the total energies. Surface structures were modeled as periodic slabs separated by approximately 25\,{\AA} of vacuum in the direction of the surface normal; a ($3\times 3$) Monkhorst-Pack $k$-point grid was found to be sufficiently accurate (convergence up to a few tens of meV) for sampling the Brillouin zone of the supercell. Convergence was considered to have occurred for a maximum energy difference of $10^{-4}$\,eV between electronic iterations and a maximum remaining force of 0.01\,eV/{\AA} per atom for ionic relaxation, respectively.

Harmonic molecular vibrations were found by means of normal mode analysis, \textit{i.e.}\ by diagonalization of the system's dynamical matrix, also including all degrees of freedom of the uppermost substrate layers (details see below); numerical derivatives of the energy with respect to the nuclear coordinates were evaluated using centered finite differences.

\subsubsection*{Surface Model\label{surf_mod}}

\begin{figure*}
  \centering
  \includegraphics[width=.75\textwidth]{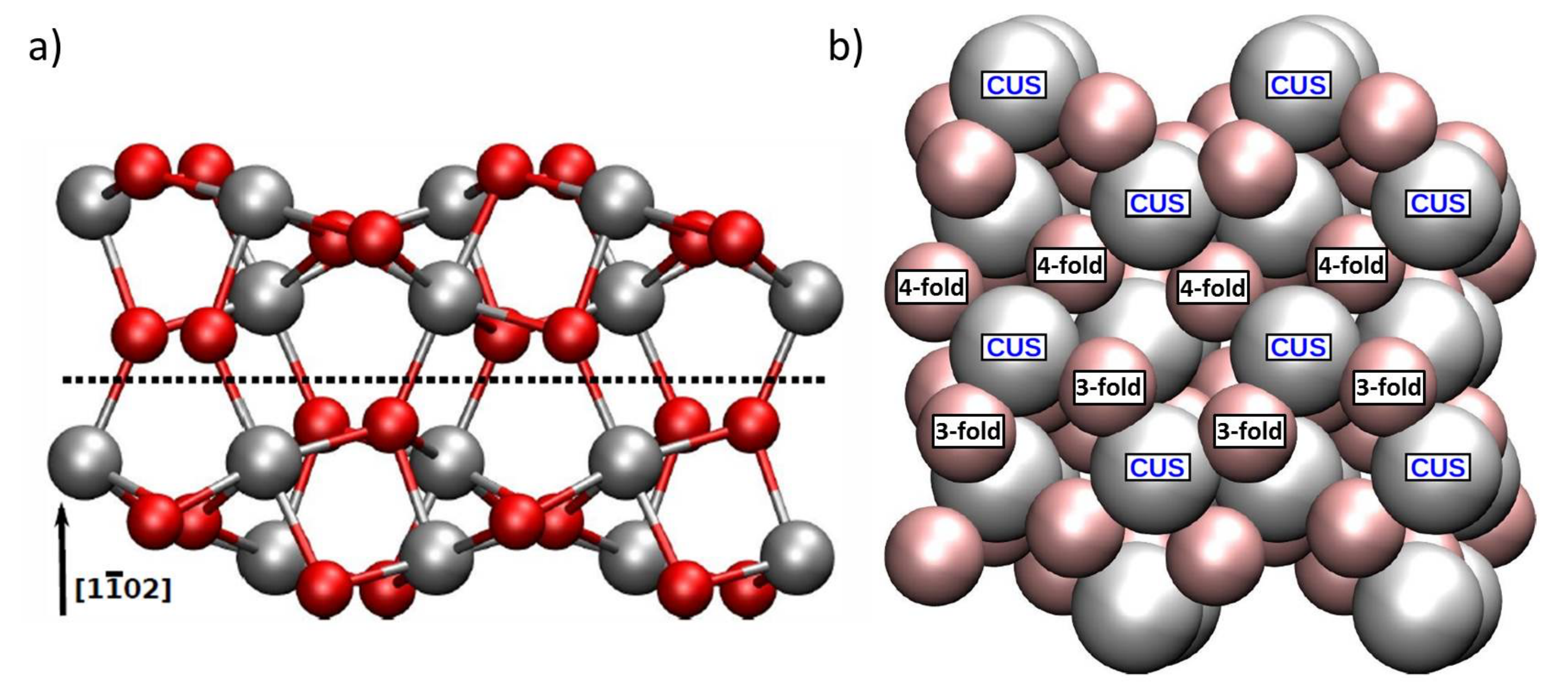}
  \caption{($2\times 2$) supercell model of the oxygen terminated $\alpha$-Al$_{2}$O$_{3}$(1\=102) surface used in this study; oxygen atoms are shown as smaller red, aluminum atoms as larger grey balls. a)~Side view (ball-and-stick representation): For further optimizations and vibrational analyses featuring adsorbate species on the top side of the slab atoms below the dotted line were kept fixed at their position initially optimized for the bare surface. b)~Top view (``van-der-Waals spheres''): Aluminum positions for water adsorption are indicated with ``CUS'' labels. 1 rows of 3-fold and one of 4-fold coordinated oxygens are labelled for reference.}
  \label{1102_surf}
\end{figure*}

As noted in detail by Trainor \textit{et al} \cite{Trainor02}, terminating the bulk unit cell of $\alpha$-Al$_{2}$O$_{3}$ along the (1\=102) plane gives a surface that cannot be repeated by translation along a lattice vector normal to the surface. Because such translation is required for the periodic boundary conditions of our calculation we follow Trainor et al and instead employ a larger ($2\times 2$) supercell (strictly speaking a pseudo supercell) model \cite{Trainor02}. The resulting slab shown in Figure~\ref{1102_surf}\,a consists of six oxygen and four aluminum layers in the vertical direction, the lower half of which were kept fixed (after initial geometry optimization of the adsorbate-free structure) during optimizations, single point energy calculations and vibrational analyses in the presence of adsorbate species on the top side of the slab. The rectangular cell correctly describing the lateral translational symmetry of the surface structure is defined by two vectors in the directions of the crystallographic [11\=20] and [\=1101] axes (see Figure~\ref{1102_surf}\,b). The calculated corresponding lattice constants were $a=9.60$\,{\AA} and $b=10.22$\,{\AA}, and deviated less than 1\% from experimental values~\cite{Trainor02}. Consistent with prior work by Kurita et al \cite{kuri10}, vertical relaxation for $\alpha$-Al$_{2}$O$_{3}$(1\=102) was found to be less pronounced than in case of the Al terminated (0001). In what follows we describe the surface Al atoms as Coordinatively Unsaturated Sites (CUS). All such CUS sites for the (1\=102) surface exhibit the same, albeit rotated, chemical environment: they are each surrounded by 2 three-fold coordinated and 2 four-fold coordinated oxygens.

\subsubsection*{Thermodynamics and Kinetics}

Water adsorption energies were evaluated as the difference
\begin{equation}
  \Delta E_\mathrm{ads} = E(\mathrm{H_2O + surface}) - \left[E(\mathrm{surface}) + E(\mathrm{H_2O})\right]
  \label{Eads}
\end{equation}
with $E$ denoting the sum of electronic energy and nuclear repulsion. Free energies $G(T)=E+H(T)-TS(T)$ were calculated considering vibrational, rotational and translational finite-temperature contributions to enthalpy, $H(T)$, and entropy, $S(T)$, in case of gas phase water (assuming $p=1$\,bar). For adsorbed species only vibrational contributions were included since rotational and translational motions are frustrated. The individual contributions were calculated according to the procedure outlined in our previous study~\cite{Wirth12} as well as in standard text books~\cite{Jensen07}. To ensure error cancellation, equal computational settings were applied to all species involved.

Surface processes such as water dissociation or diffusion were studied using the Nudged Elastic Band (NEB) method (see Ref.~\citenum{Jonsson98} and references therein) as implemented in the modified VASP version of J\'onsson and co-workers. This includes an improved estimate to the band tangent~\cite{Henkelman00a} as well as a climbing image (CIS) scheme~\cite{Henkelman00b} for locating the transition state along the Minimum Energy Path (MEP) of the reaction. The NEB procedure used here includes successive linear interpolation steps as described in Ref.~\citenum{Wirth12}, ending up with eleven images (plus reactant and product image) along the path. Transition states were characterized by a single imaginary frequency corresponding to a vibrational mode that is the coordinate connecting the reactant and product states. Reaction rates were estimated as a function of temperature using the expression\cite{Eyring35},
\begin{equation}
  k(T) = \frac{k_\mathrm{B}T}{h}e^{-\Delta G^\ddagger(T)/(k_\mathrm{B}T)}
  \label{Eyring}
\end{equation}
where $k_\mathrm{B}$ is the Boltzmann constant, $h$ the Planck constant and the activation free energy $\Delta G^\ddagger = G^\ddagger - G_\mathrm{R}$ is the difference between the free energy of the reactant and transition states. Note that, because we employ \emph{standard} DFT approaches there is no difference in the calculated energies of OD and OH fragments. We account for the effect of increased mass in vibrational properties and free energies by scaling the hydrogen mass from 1u to 2u. Additionally, since we only deal with D$_2$O, and its fragments, non-classical corrections to equation \ref{Eyring} are neglected.


\section{Results and Discussion\label{res_disc}}

\subsection{Experimental UHV Results\label{res_exp}}

As described above, we expect our \ce{D2O} dosing procedure to result in creation of a multilayer of ice. To verify the presence of these multilayers, and their subsequent removal with sample thermal annealing, we performed TPD measurements one example of which is shown in Figure \ref{SFG_data1}a. The TPD spectrum is clearly dominated by a double peak structure with a long tail to the high temperature side. We and others have previously observed this long tail at higher temperatures in TPD spectra of water adsorption on  $\alpha$-Al$_{2}$O$_{3}(0001)$ \cite{elam98,kirsch2014} and assigned it to the recombinative desorption of dissociatively adsorbed water. We follow this assignment here for water adsorption on the $\alpha$-\ce{Al2O3}(1\=102) surface. In addition to the tail, two desorption maxima, centered at 162 and 175 K are apparent (see Figure \ref{SFG_data1}a). Because the 162 K peak increases in magnitude with increasing dosing time while the 175 K does not, we assign the former to the desorption of multilayer water (\textit{i.e.}\ ice) and the later to the desorption of surface ($<$ 1 monolayer) molecular water.

\begin{figure}
\begin{center} 
\includegraphics[width=0.5\textwidth]{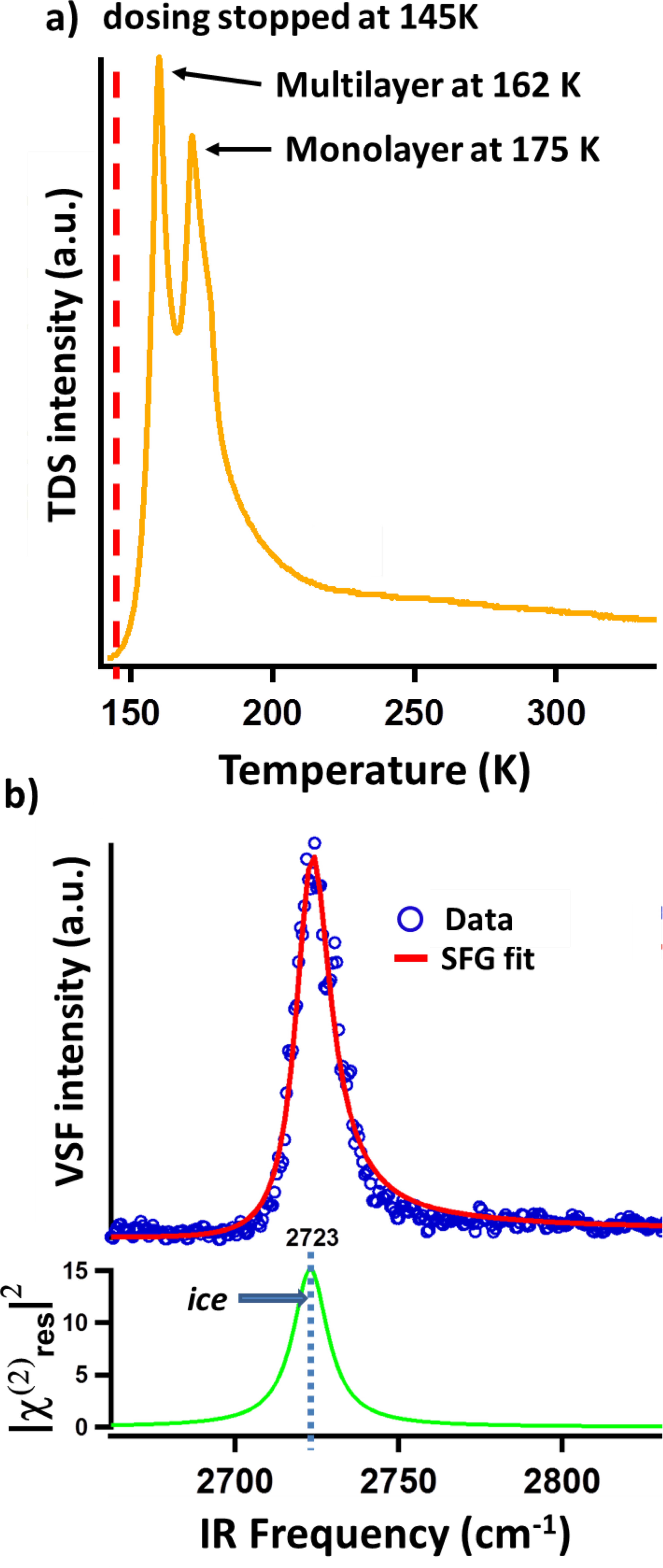}
\end{center} 
\caption{a) the TPD spectrum of mass 20 (\ce{D2O}) for a sample, prepared as described in the text. To perform the TPD, after stopping \ce{D2O} dosing at 145 K we applied a heating ramp of 100 K/min and detected the desorption of mass 20. The desorption maximum at 162 K is assigned to water desorbing from the the weakly bond multilayer of ice (it can be increased in intensity by dosing for longer at low surface temperature). The peak at 175 K is assigned to monolayer desorption. Its amplitude is independent of low temperature dosing time. b) VSF spectrum of $\alpha$-\ce{Al2O3}(1\=102) surface covered by \ce{D2O} ice and showing the characteristic free-OD peak of the ice/vacuum interface \cite{kirsch2014}.}
\label{SFG_data1}
\end{figure}

Performing VSF measurements of samples prepared as described above with no subsequent thermal annealing results in a VSF spectrum with a strong single resonance (see Figure \ref{SFG_data1}b). Using the line shape model discussed above, fitting the data allows us to extract a resonance center frequency of 2723 cm$^{-1}$ for this feature as expected for the free-OD of \ce{D2O} ice \cite{kirsch2014}. 

If our assignment of the TPD data is correct, rapid thermal annealing to temperatures above 175 K should result in a surface that has sub-monolayer \ce{D2O} coverages. Given that the spectral response shown in Figure \ref{SFG_data1}b is that of the the ice surface one might expect that after evaporating all ice there would be a notable decrease in intensity of the VSF spectrum (the monolayer coverage of adsorbed water is disordered relative to the ice surface) and a change in lineshape (water's local hydrogen bonding environment also changes with respect to that found in ice). As shown in Figure \ref{SFG_data2}a, our measurements confirm these expectations. Clearly, after annealing to 185 K what was a single intense resonance centered at 2723 cm$^{\text{-1}}$ is now two weaker resonances centered at 2733 and 2772 cm$^{\text{-1}}$. Scanning our IR source over the range 1500 - 3000 cm$^{\text{-1}}$ and at other polarizations gave no other detectable resonant features.

Because the TPD data (see Figure \ref{SFG_data1}a) make clear that at 185 K our $\alpha$-\ce{Al2O3}(1\=102) surface has only submonolayer coverages of D$_{\text{2}}$O, it seems reasonable to suggest that the two resonances apparent after annealing to 185 K at 2733 and 2772 cm$^{\text{-1}}$ (see Figure \ref{SFG_data2}a) must be the result of either dissociatively or molecularly adsorbed water. Prior studies have found that some hydroxyl groups resulting from dissociative adsorption of \ce{H2O} on $\alpha$-\ce{Al2O3}(1\=102) are stable in UHV up to   temperatures of $\approx$ 700 K \cite{schildbach1993}. This finding suggests that if an observed OD stretch resonance originates from molecularly adsorbed \ce{D2O}, heating to temperatures $>$ 300 K should result in its disappearance while OD stretch modes related to dissociatively adsorbed \ce{D2O} should still be present. 

Figure \ref{SFG_data2}b shows VSF spectra spectra collected from samples annealed to temperatures above 300 K and rapidly cooled to 135 K for measurement. Clearly the 2733 cm$^{\text{-1}}$ feature apparent at 185 K is now absent (see Figure \ref{SFG_data2}a) and the intensity of the resonance centred at 2773 cm$^{-1}$ decreases with increasing maximum temperature.

\begin{figure}
\begin{center} 
\includegraphics[width=0.57\textwidth]{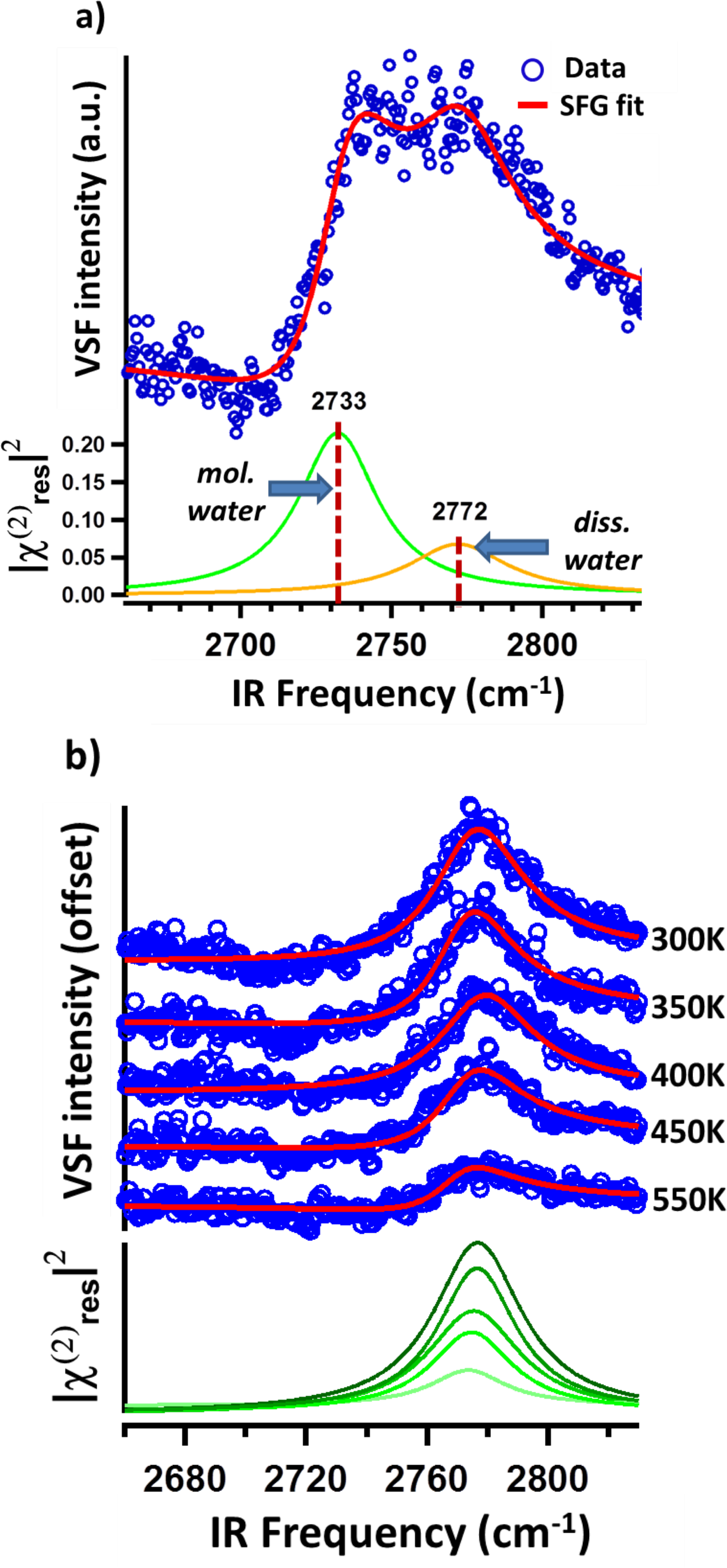}
\end{center} 
\caption{VSF spectra of interfacial water at different surface temperatures. a) OD stretch spectra of a sample rapidly annealed to 185 K and immediately cooled to 135 K for characterization. The coverage is in the submonolayer regime. The spectrum clearly shows two features the lower frequency of which is assigned to molecularly, the higher frequency of which to dissociatively, adsorbed \ce{D2O} (see text for discussion). b) Series of VSF spectra collected from a $\alpha$-\ce{Al2O3}(1\=102) rapidly annealed to the indicated temperature. Between each annealing step the sample is cooled to 135 K for VSF spectra characterization. Clearly only a single feature is apparent whose intensity decreases with annealing temperature. This feature is assigned to (an) OD stretch mode of dissociatively adsorbed \ce{D2O}}
\label{SFG_data2}
\end{figure}

This data is thus consistent with a scenario in which the 2733 cm$^{\text{-1}}$ resonance is an OD stretch of molecular \ce{D2O} while the feature centered at 2773 cm$^{\text{-1}}$ is the result of dissociative adsorption. In this scenario, and consistent with the TPD data in Figure \ref{SFG_data1}a, the decrease in intensity of the 2773 cm$^{\text{-1}}$ feature with increasing temperature is the result of decreasing water coverage with increasing temperature. While consistent with experiment, clearly this understanding of our observed trends still leaves questions: (i) What are the thermodynamics and kinetics that control \ce{D2O}/$\alpha$-\ce{Al2O3}(1\=102) interaction? (ii) Relatedly, why do we observe only one spectral feature associated with dissociatively adsorbed \ce{D2O} (one might expect at least two per energetically plausible dissociation mechanism)? (iii) Why do we observe only one spectral feature associated with molecularly adsorbed \ce{D2O}? To help answer these questions we turn to theory.   


\subsection{Computational Results\label{res_theo}}

\subsubsection*{Water Adsorption}
As can be seen from the model introduced in Section~\ref{ss:comp}, the surface density of CUS positions for water adsorption on the $\alpha$-Al$_{2}$O$_{3}$(1\=102) surface is 0.08\,{\AA}$^{-2}$, 60 \% higher than the 0.05\,{\AA}$^{-2}$ for the (0001) surface~\cite{Wirth12}. To connect to our sub-monolayer experimental results, we here focus on the description of water adsorption, dissociation and diffusion on the $\alpha$-\ce{Al2O3}(1\=102) surface in the low coverage limit. To our knowledge there is no prior work that has previously explored water reactivity on this surface at these coverages. By placing a single \ce{D2O} molecule in the ($2\times 2$) supercell, a surface with 1/8 of a monolayer with respect to the CUS positions results.

\begin{figure*}
  \centering
  \includegraphics[width=.8\textwidth]{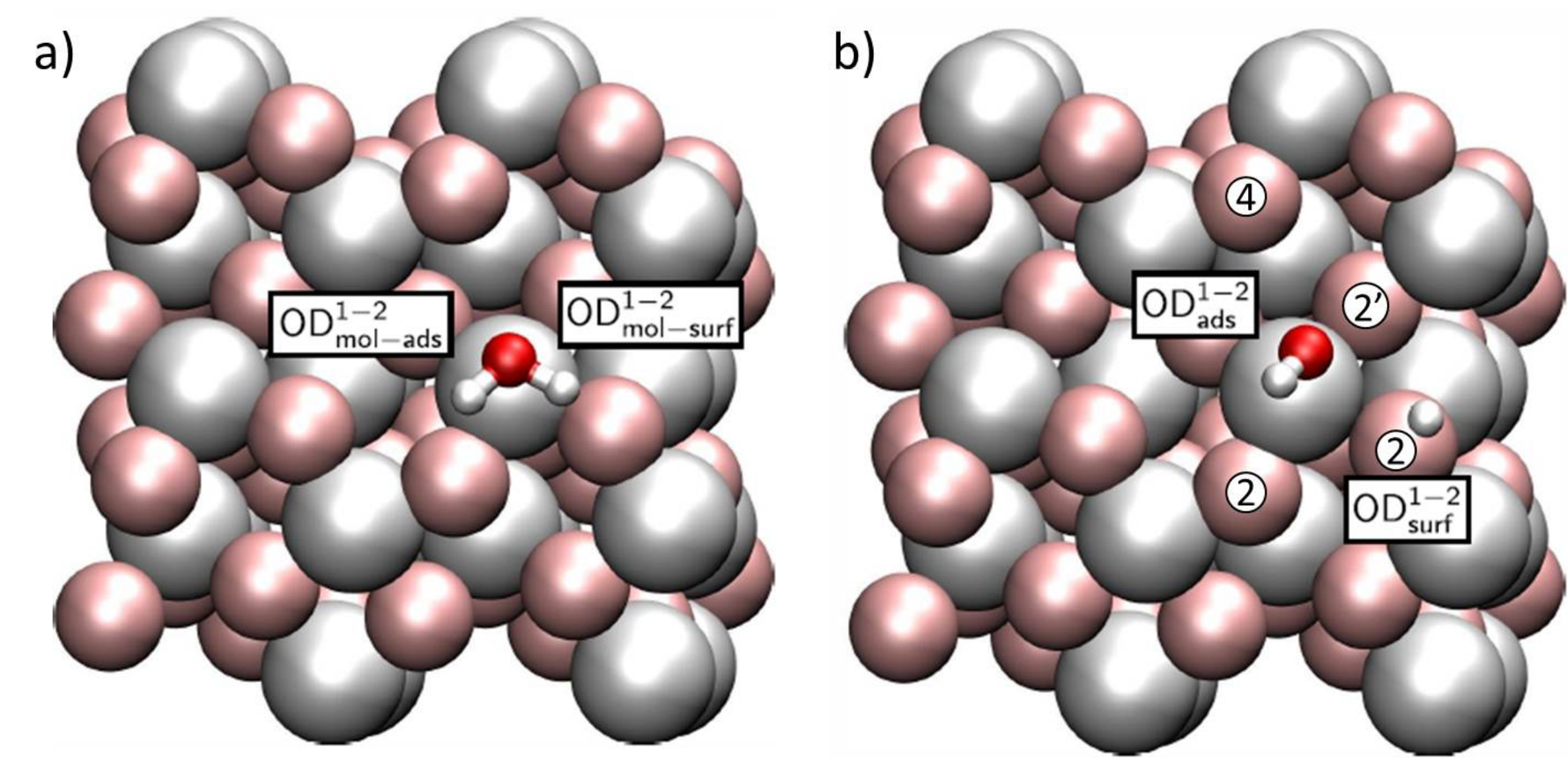}
  \caption{Adsorption geometries for a) molecular and b) dissociated water in the 1-2 configuration. For convenience, surface atoms are shown as ``van-der-Waals spheres'' in pale colours, adsorbate species in a ball-and-stick representation. Possible adsorption sites are also designated in b).}
  \label{1-2-geoms}
\end{figure*}

Because the two aluminum CUS atoms per unit cell of the $\alpha$-Al$_{2}$O$_{3}$(1\=102) surface structure have the same, albeit rotated, chemical environment (see Figure \ref{1102_surf}\,b), only two distinct water adsorption geometries are possible. In the first, shown in Figure \ref{1-2-geoms} the oxygen of an adsorbing \ce{D2O} sits atop a surface Al. This \ce{D2O} either adsorbs dissociatively, in which case the OD originating from \ce{D2O} remains on the surface Al and the extra D is adsorbed by the nearest neighbor three-fold coordinated surface oxygen, labelled a \textit{2} site, or molecularly, in which case it does not. Note that for molecularly adsorbed \ce{D2O} in this configuration the two OD groups are not equivalent: the OD angles with respect to the surface normal and bond lengths both differ (see Table \ref{E_G_ads} and Supplementary Information for further structural information). In what follows we refer to the four OD groups shown in Figures \ref{1-2-geoms}a and \ref{1-2-geoms}b  using the superscript $^{\text{1-2}}$. If this molecule dissociatively adsorbs it forms an OD group originating from \ce{D2O}, denoted by the subscript \textit{ads} or from bond formation between a surface oxygen and deuterium from \ce{D2O}, denoted by subscript \textit{surf}. To refer to the OD in molecularly adsorbed \ce{D2O} that is OD$_{\text{ads}}$ if the molecule dissociates we employ the subscript \textit{mol-ads} and to refer to the D from \ce{D2O} that forms a bond with surface oxygen when \ce{D2O} dissociates we use the subscript \textit{mol-surf}. As shown in Figure \ref{1-4-geoms}, in the second possible set of adsorption geometries an impinging \ce{D2O} molecule adsorbs on the flank of a surface Al and, when it dissociates, the surface D adsorbs on the next nearest neighbour 3-fold coordinated oxygen (\textit{i.e.}\ the \textit{4} site, see Figure \ref{1-2-geoms}b for definition of site labelling convention). The notation used to describe the four possible ODs in this situation is the same as above except a superscript \textit{1-4} is employed. Adsorption of D on 4-fold coordinated oxygens, the $2^{\prime}$ sites, is unfavorable.

\begin{figure*}
  \centering
  \includegraphics[width=.8\textwidth]{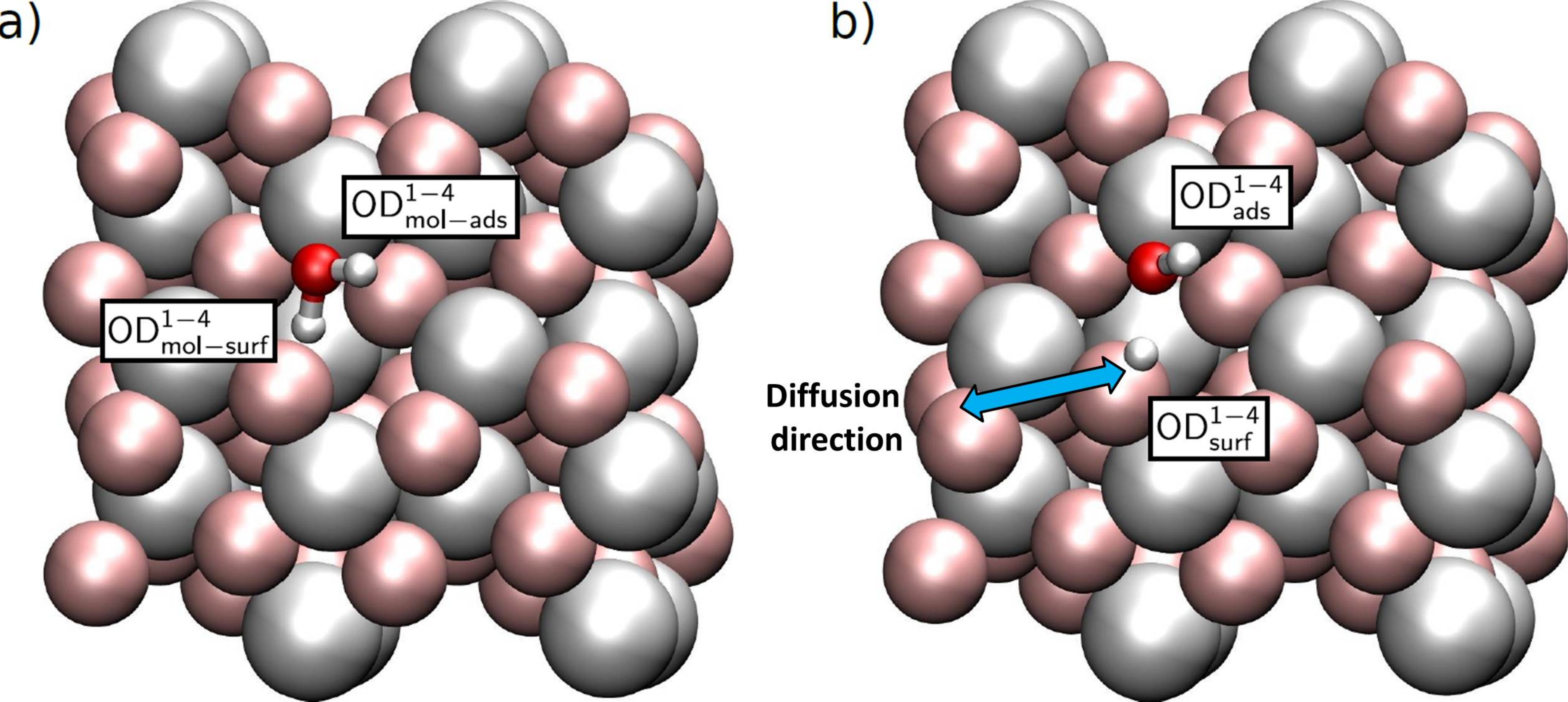}
  \caption{Adsorption geometries for a) molecular and b) dissociated water in the 1-4 configuration. The blue arrow indicates a portion of the \textit{diffusion channel} of the deuteron discussed in the text.}
  \label{1-4-geoms}
\end{figure*}

Calculated energies and free energies of each of these configurations are shown in Table~\ref{E_G_ads} along with the angles of each OD group relative to the surface normal. These results suggest that, perhaps counter intuitively, both 1-4 geometries are favoured over the 1-2 geometries by approximately 0.4 eV. This is presumably due to the coordination environment of the CUS atom being tilted with respect to the surface normal in the 1-4 allowing for a stabilizing hydrogen-bond like interaction between the water D and surface O atom (for the molecular structure) or deuteroxyl O atom and the newly formed surface OD group (for the dissociated structure) close to the opposing CUS (see Figure~\ref{1-4-geoms}). For both the 1-2 and 1-4 adsorption geometries we note further that the molecular and dissociated forms are  $\approx$ energetically degenerate: within 0.06 eV. We have shown previously that such degeneracy does not occur for the analogous 1-2 water adsorption configuration, but does for the 1-4, on the $\alpha$-\ce{Al2O3}(0001) surface \cite{Wirth12,kirsch2014}. As we will see below these differences lead to dramatically different surface reactivity. 

\begin{table}
  \centering
  \begin{tabular}{lccc}
    \toprule
    \cmidrule(r){2-3}\cmidrule(r){4-4}
    \textsc{Configuration}	&	$E_{\mathrm{ads}}$	&	$G_{\mathrm{ads}}(T=0\,\mathrm{K})$	&	$G_{\mathrm{ads}}(T=135\,\mathrm{K})$	\\
    \midrule
    1-2-mol.		&	$-1.07$			&	$-1.02$					&	$-0.83$					\\
    1-2-diss.		&	$-1.11$			&	$-1.08$					&	$-0.89$					\\
    \midrule
    1-4-mol.		&	$-1.48$			&       $-1.45$                   		&	$-1.25$					\\
    1-4-diss.		&	$-1.53$			&	$-1.51$					&	$-1.31$					\\
    \bottomrule
  \end{tabular}
 \caption{Calculated (free) energies of adsorption (in eV) for low-coverage molecular (mol.) and dissociated (diss.) heavy water (D$_2$O) on the $\alpha$-Al$_{\text{2}}$O$_{\text{3}}$(1$\bar{\text{1}}$02) surface.}
  \label{E_G_ads}
\end{table}

\subsubsection*{Surface Diffusion}
Presumably if either OD$_{\text{ads}}$ or OD$_{\text{surf}}$  (see Figures \ref{1-2-geoms} and \ref{1-4-geoms} for OD fragment classification) could diffuse this would both be important for surface reactivity and lead to possible changes in ensemble averaged OD stretch spectral response that might be apparent in our measurements. 

Much prior work and preliminary studies here strongly suggest that the OD fragment resulting from \ce{D2O} dissociative adsorption, \textit{i.e.} OD$_{\text{ads}}$, forms a covalent bond to a surface aluminum that is sufficiently strong to be fixed on all relevant timescales \cite{Wirth12,kirsch2014}. As a result we here investigate the likelihood of only two types of surface diffusion: i)~diffusion of a water molecule between neighbouring CUS positions each adsorbed in the low energy 1-4 configuration and ii)~migration of the detached water deuteron from the 1-4 position to a neighbouring surface oxygen atom still further away. Because simple thermodynamical considerations dictate that the 1-4 configuration is 10$^{7}$ more probable than the 1-2 (given $\text{G}_{\text{1-4}}- \text{G}_{\text{1-2}}=\Delta G= -0.4$\,eV at $T=300$\,K, according to Boltzmann statistics $e^{-\Delta G/(k_\mathrm{B}T)}\approx 10^{7}$ more probable), we did not consider diffusion of the deuteron from the 1-4 to 1-2 positions.

As discussed in the Methods section, we address each of these processes by calculating the relative free energies of the reactant and product states and the reaction coordinate that connects them sampled via the nudged elastic band method.
\begin{figure}
  \centering
  \includegraphics[width=.8\textwidth]{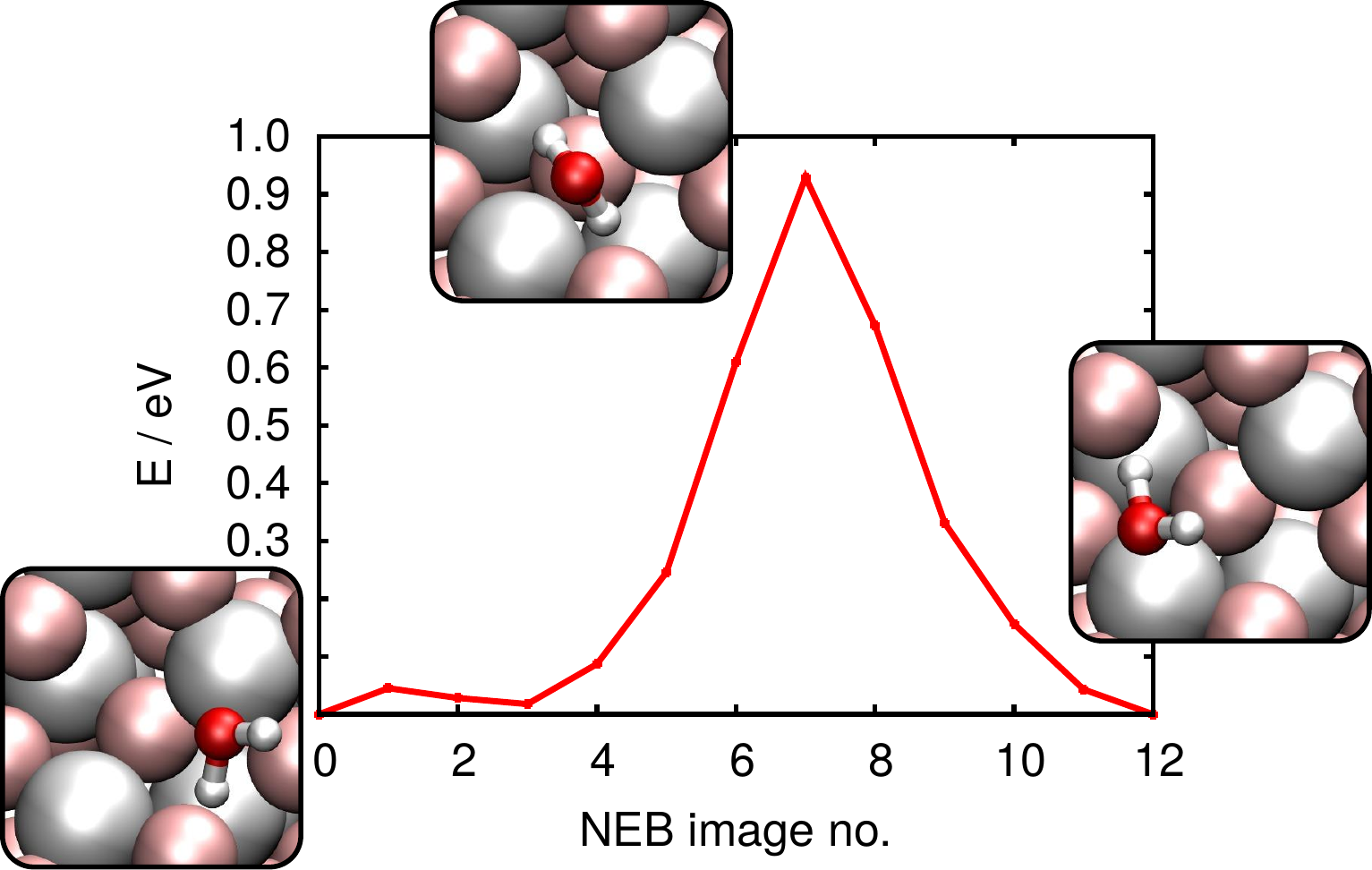}
  \caption{NEB-derived energy profile for molecular water diffusion between two neighboring CUS positions (adsorption in 1-4 orientation each); see insets for reactant, product and transition state geometries. The diffusion barrier is well below the adsorption energy but too high for the process to be relevant under low to moderate temperature conditions.}
  \label{mol_diff}
\end{figure}
The NEB-derived energy profile for diffusion of a water molecule between two neighboring 1-4 CUS positions is shown in Figure~\ref{mol_diff}. The process proceeds \emph{via} rotation of the water molecule through a rather high energy transition state: 0.93\,eV. According to Equation~\ref{Eyring} at 300\,K this results in a rate constant of $2\times 10^{-3}$\,s$^{-1}$. At the VSF measurement condition of $T=135$\,K this rate drops to $9\times 10^{-23}$\,s$^{-1}$ suggesting diffusion of molecular water under VSF measurement conditions does not occur. While diffusion between 1-4 configurations would be possible during high temperature annealing clearly such changes do not influence surface energetics, in the low coverage limit all 1-4 configurations are the same.
\begin{figure}
  \centering
  \includegraphics[width=.8\textwidth]{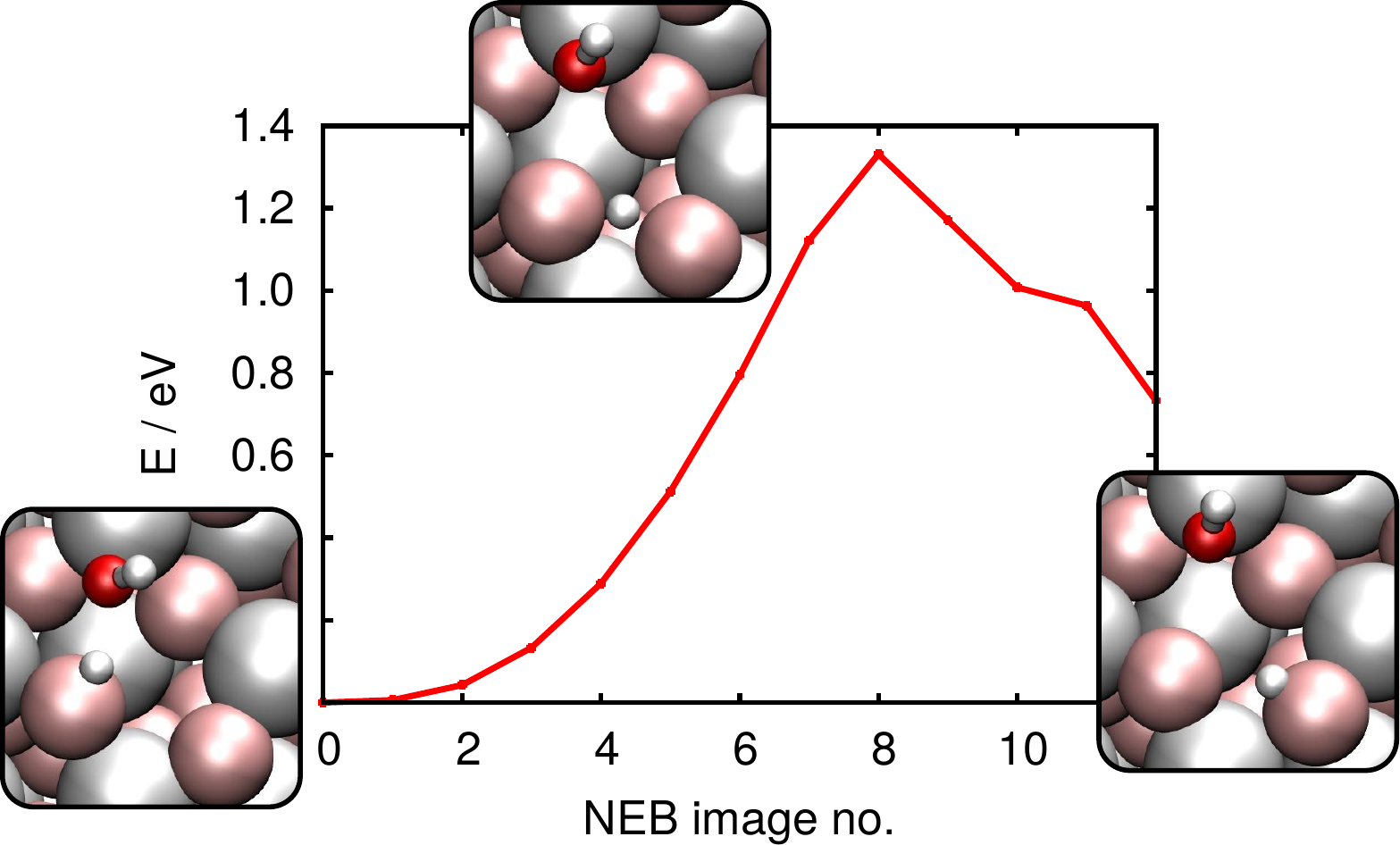}
  \caption{NEB-derived energy profile for diffusion of a proton/deuteron from the 4 position to a neighboring surface oxygen atom; see insets for reactant, product and transition state geometries.}
  \label{H_diff}
\end{figure}

As discussed above, in either the 1-2 or 1-4 adsorption configuration, the dissociated D forms a covalent bond with a 3-fold coordinated surface oxygen. On the (1\=102) surface such oxygens form horizontal rows in Figures ~\ref{1-2-geoms} and \ref{1-4-geoms} (\textit{i.e.} in the crystallographic (11\=20) direction) that are separated by 4-fold coordinated oxygens (\textit{i.e.} in the crystallographic (\=1101) direction). Because D bond formation with such saturated oxygens is extremely unfavorable, surface structure suggests that D diffusion, if it occurs, should take place along \textit{diffusion channels} (as shown in Figure \ref{1-4-geoms}). Because, as discussed above, the 1-2 adsorption configuration should exist in \emph{extremely} small numbers, we here only consider the diffusion of the deuteron from the 4 position in the 1-4 dissociated configuration (Figure~\ref{1-4-geoms}\,b)) to its nearest neighbour surface oxygen atom. From the resulting NEB-derived energy profile shown in Figure~\ref{H_diff} clarifies that such motion is thermodynamically unfavourable, by 0.73 eV and has a barrier of 1.33\,eV. In order to assess the thermodynamic favourability of other product states, a number of different configurations with distances between OD group and deuteron larger than in the 1-2 or 1-4 dissociated geometry were also optimized. No configuration was found in which the new position of deuteron was less than 0.5\,eV destabilised, with respect to the initial, stable 1-4 dissociated geometry.

Armed with free energies at each point along the reaction coordinate the rate constant for proton/deuteron ``escape'' from the 1-4 dissociated configuration is, at 135 K, $4\times 10^{-34}$\,s$^{-1}$, at 300 K, $3\times10^{-8}$\,s$^{-1}$ and at 550 K, $70$\,s$^{-1}$. This suggests that deuteron diffusion is essentially impossible at the temperatures of VSF analysis and for samples rapidly annealed to 185 K (\textit{i.e.}\ Figure \ref{SFG_data2}a) but that it is significant for samples annealed to temperatures 300-550 K (\textit{i.e.}\ Figure \ref{SFG_data2}b). However, given the 1-D nature of deuteron diffusion and the structural similarity of all 3-fold coordinated surface oxygens, it seems clear that thermodynamic considerations suggest the great majority of deuterons are in the 1-4 configurations (\textit{i.e.}\ given a destabilization energy of $\Delta G = -0.5$ eV, at 300 K 1-4 configurations are  $e^{-\Delta G/(k_\mathrm{B}T)}\approx 10^{9}$ more probable). Taken together these calculations suggest that at temperatures from 135-185 K deuteron diffusion does not happen on the timescale of the measurements. While diffusion does occur from 300-550 K, thermodynamics suggests that deuteron locations other than those shown in Figure \ref{1-4-geoms} are exceptionally uncommon.

\subsubsection*{Water Dissociation and its Vibrational Fingerprint}

Having understood the energetics of each adsorption configuration and the possible influence of surface diffusion, we next turn our attention to the characteristic frequency of each surface species. We do so by conducting normal mode analyses for the four possible structures, and eight possible OD groups, shown in Figures~\ref{1-2-geoms} and \ref{1-4-geoms}. 
\begin{table}
  \centering
  \begin{tabular}{lccc}
    \toprule
					&	\multicolumn{2}{c}{\textsc{Theory}}			&	\textsc{Experiment}		\\
    \textsc{Assignment}				&	$\theta$(deg)	&	freq (cm$^{\text{-1}}$)	&	freq (cm$^{\text{-1}}$)	\\ 
    \midrule
    OD$^{\text{1-2}}_{\mathrm{ads}}$		&	41		&	2782 			&	---			\\[0.03in]
    OD$^{\text{1-4}}_{\mathrm{ads}}$  		&	34		&	2779 			&	2772			\\[0.03in] 
    OD$^{\text{1-4}}_{\mathrm{mol-ads}}$	&	49		&	2737			&	2733			\\[0.03in]
    OD$^{\text{1-2}}_{\mathrm{mol-ads}}$	&	67		&	2688			&	---			\\[0.03in]
    OD$^{\text{1-2}}_{\mathrm{surf}}$  	&	37		&	2680			&	---			\\[0.03in]
    OD$^{\text{1-2}}_{\mathrm{mol-surf}}$	&	84		&	2493			&	---			\\[0.03in]
    OD$^{\text{1-4}}_{\mathrm{surf}}$  	&	50		&	1958			&	---			\\[0.03in]
    OD$^{\text{1-4}}_{\mathrm{mol-surf}}$	&	112		&	1651			&	---			\\
    \bottomrule
  \end{tabular}
\caption{Calculated OD fragment orientations, calculated normal mode frequencies and vibrational frequencies extracted from our VSF data.}  
  \label{tab_freqs}
\end{table}
The resulting eight frequencies are given in Table~\ref{tab_freqs} together with the associated angles between OD bond vector and surface normal. From these values an almost quantitative agreement of the calculated OD$^{1-4}_\mathrm{ads}$ and OD$^{1-4}_\mathrm{mol-ads}$ frequencies with the experimental resonances is clear. 

\begin{figure}
  \centering
  \includegraphics[width=.8\textwidth]{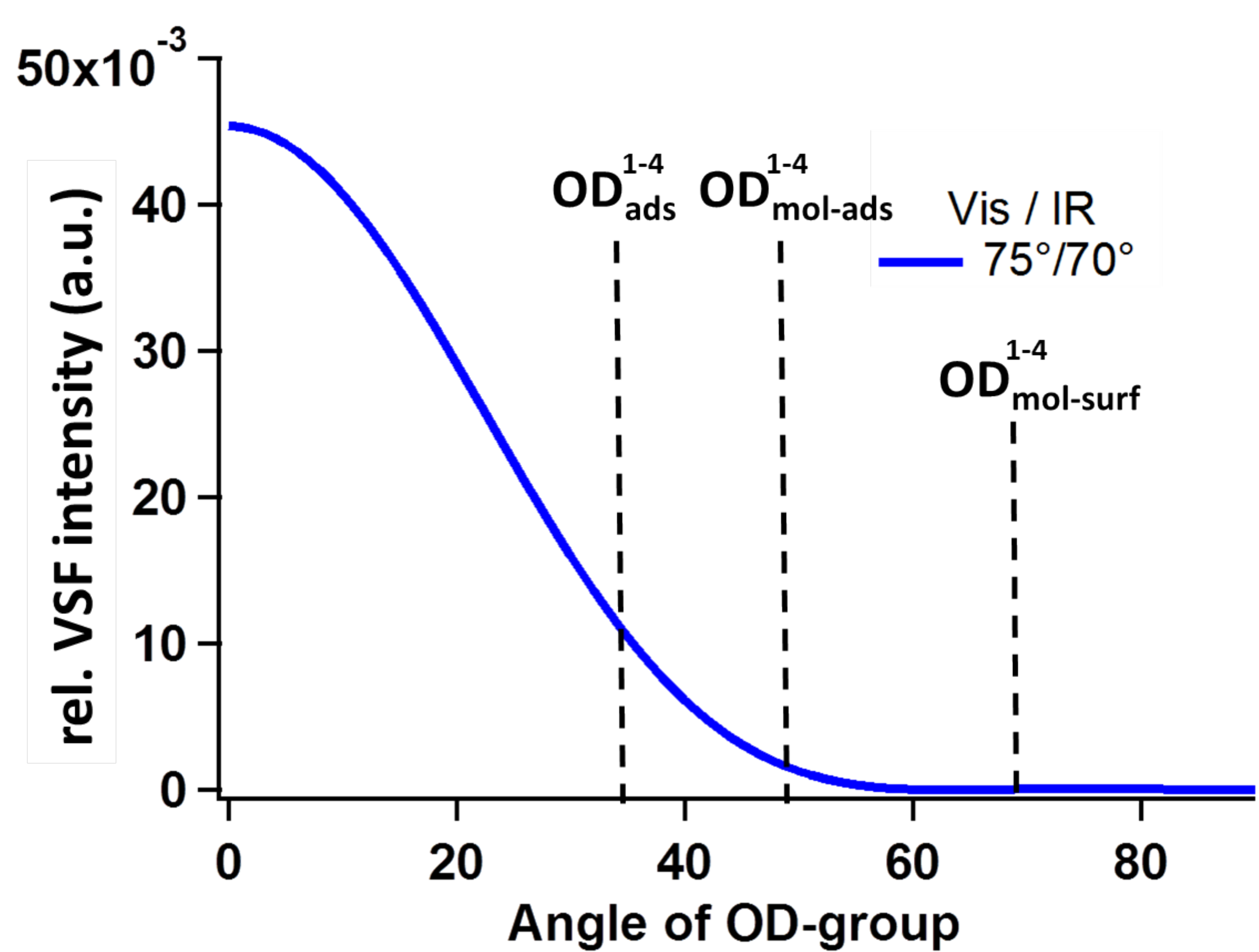}
  \caption{Estimated intensity of the VSF response (using the $ppp$ polarization scheme) as a function of OD orientation given by the angle between OD bond vector and surface normal; details of the underlying model are outlined in the Supporting Information. The angle for the OD$^{1-4}_{\mathrm{mol-surf}}$ is drawn in this scheme at 68 degree, even though given in Table \ref{tab_freqs} as 112 degree, because the measured $I_{\text{VSF}}$ signal is insensitive to changes in OD orientation of 180$^{\circ}$.}
  \label{orient}
\end{figure}

Considering the well-known GGA-typical overestimation of fundamental frequencies, the nearly perfect overlap between measured and calculated frequencies is likely the result of a fortuitous cancellation between  DFT error and anharmonicity. However, much prior work shows that \textit{relative} harmonic frequencies do not typically suffer from either effect and clearly restating the results in Table~\ref{tab_freqs} in these terms would lead to the same assignment \cite{kirsch2014}. Given this assignment, however, we are still left with the obvious question: why are only two of the possible eight modes apparent in experiment?

Assuming the ratio of 1-2 and 1-4 adsorbed \ce{D2O} molecules reflects equilibria between all species, the results in Table \ref{E_G_ads} clearly suggest that the measurable OD stretch spectral response should be dominated by \ce{D2O} molecules in the 1-4 configuration -- as noted above they should be 10$^{\text{7}}$ more abundant than 1-2 structures. However, this still leaves four possible species, OD$^{\text{1-4}}_{\text{ads}}$, OD$^{\text{1-4}}_{\text{mol-ads}}$, OD$^{\text{1-4}}_{\text{surf}}$ and OD$^{\text{1-4}}_{\text{mol-surf}}$, of which only the first two seem to appear in our measurement. Because a \ce{D2O} molecule in the 1-4 configuration must be \emph{either} molecularly \emph{or} dissociatively adsorbed, for every OD$^{\text{1-4}}_{\text{ads}}$ there must be an OD$^{\text{1-4}}_{\text{surf}}$ and for every OD$^{\text{1-4}}_{\text{mol-ads}}$ OD$^{\text{1-4}}_{\text{mol-surf}}$. As a result the absence of all such resonance features cannot be the result of thermodynamics. 

As discussed above, the measured VSF intensity, given fixed incident beam angles and polarizations, is a function of molecular orientation. Thus one possibility is that the missing resonances exist, but because of the particular combination of incident beam angles and polarizations as well as fragment orientation our VSF measurement is insensitive to their presence. Using the theory discussed earlier (and in more detail in the Supporting Information) we have calculated the expected relative VSF response for an OD group, given our experimental configuration, as a function of molecular orientation. These results are shown in Figure \ref{orient}. Clearly they suggest that, given the calculated orientation of OD$^{\text{1-4}}_{\text{mol-surf}}$, \textit{i.e.} 112(68)$^{\circ}$ with respect to the surface normal, it should not appear in our results.
\begin{figure}
  \centering
  \includegraphics[width=.7\textwidth]{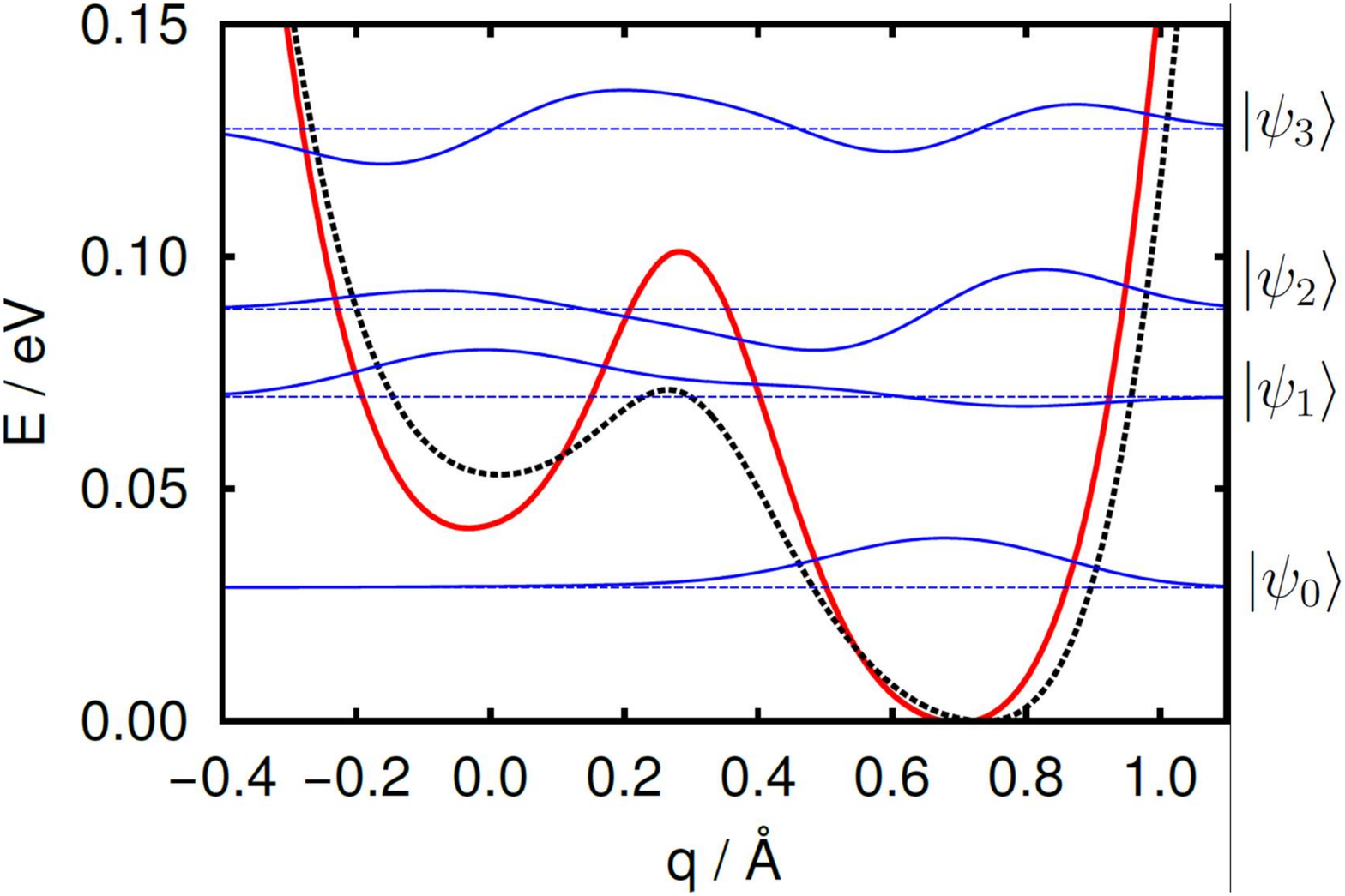}
  \caption{NEB-derived potential for 1-4 water dissociation on the $\alpha$-Al$_{2}$O$_{3}$(1\=102) surface based on pure PBE (dotted black) and HSE06 (solid red). The latter was obtained by single-point calculations on PBE geometries; both potentials were aligned with respect to the 1-4 dissociated minimum (\textit{i.e.}\ the minimum on the right). Vibrational eigenvalues (dashed blue) and wave functions (solid blue) are given for the HSE06 potential; the reaction coordinate $q$ is a projection onto the linear transit path between both minima (see text for details).}
  \label{eigenfct}
\end{figure}

This leaves only the OD$^{\text{1-4}}_{\text{surf}}$ mode that, if our assignments are correct, we would expect to observe but have not. One possible explanation for this feature's absence is that the uncoupled harmonic oscillator assumption that underlies the frequencies shown in Table \ref{tab_freqs} does not apply. One scenario in which this approximation would be expected to fail is if the dissociation barrier in the 1-4 configuration is of order one vibrational quanta (1958 cm$^{-1}$ = 0.24 eV). To evaluate this possibility we calculated the NEB-derived 1-4 dissociation potential using two different exchange-correlation functionals (see Figure~\ref{eigenfct}).

Here, the Linear Transit (LT) path between both minima was found to be a suitable coordinate for describing the dissociation reaction:
\begin{equation}
  q = \vec{x}\cdot\vec{u} = \left(\vec{R}-\vec{R}_\mathrm{mol}\right)\cdot\vec{u} \quad ,
\end{equation}
with $\vec{R}$ denoting any geometry during the course of the reaction, $\vec{R}_\mathrm{mol}$ that of molecularly adsorbed \ce{D2O}, and with $\vec{u}$ as the unit vector pointing from molecularly to dissociatively adsorbed states. Assuming a unitary transformation between the set of $N$ cartesian coordinates $\{x_i\}$ and a set of normal coordinates $\{q_i\}$, with the above $q$ solely representing the reactive movement of ions during the dissociation process and the other $N-1$ coordinates $q_i$ being largely constant, the corresponding nuclear LT-Hamiltonian reads:

\begin{equation}
  \hat{H} = -\sum_{i=1}^N{\frac{\hbar^2}{2m_i}\frac{\partial^2}{\partial x_i^2}} + V(x_1,x_2,\dots,x_N) \approx -\frac{\hbar^2}{2\mu}\frac{\mathrm{d}^2}{\mathrm{d}q^2} + V(q)
  \label{Heq}
\end{equation}

The associated mass $\mu$ is then given by the atomic masses $m_i$ and the elements $u_i$ of the coordinate unit vector:
 
\begin{equation}
  \mu = \left(\sum_{i=1}^N{\frac{u_i^2}{m_i}}\right)^{-1}\!= 2.33\,\mathrm{amu}
\end{equation}

The potential $V(q)$ between molecular and dissociated minimum was again derived using a NEB procedure and extrapolated on both sides to obtain a double-well shape (dotted black line in Figure~\ref{eigenfct}); further analysis, however, was performed for a potential shape modified according to single-point calculations for the PBE-based geometries but using the HSE06 hybrid functional (solid red line in Figure~\ref{eigenfct}) in order to improve on the description of the transition state region. The resulting dissociation barrier is only 0.06 eV at the HSE06 level of theory, about three times lower than on the (0001) surface \cite{Wirth12}. 

Given this dissociation potential, we found numerical solutions to the vibrational Schr\"odinger equation by diagonalization of the Hamiltonian according to Equation~\ref{Heq} in a sinc-DVR basis~\cite{Colbert92} (500 grid points on the range $[-0.53,1.22]$\,{\AA}). Details of the whole procedure are outlined in Ref.~\citenum{Wirth12} (with the only difference that instead of analytic fitting in the present work interpolation of data points was done using cubic splines). The resulting vibrational eigenvalues and wavefunctions are shown in Figure~\ref{eigenfct} and clarify why no harmonic OD$^{1-4}_{\mathrm{surf}}$ signal can be found in experiment: while the system's ground state is localized, as expected, on the dissociated (right-hand) side of the potential, the first excited state is preferentially molecular in character (localized on the left-hand side of the potential) and the second excited state, has, once again more dissociated. Since the OD$^{1-4}_{\mathrm{surf}}$ mode largely corresponds to a change of the system geometry along coordinate $q$, the calculated eigenvalues strongly suggest the OD stretch vibration of OD$^{1-4}_{\mathrm{surf}}$ cannot be described in the normal-mode picture with a frequency of 1958\,cm$^{-1}$ (see Table~\ref{tab_freqs}). Instead, it is a transition between a vibrational ground and second excited state, with a frequency of $\approx$ 484\,cm$^{-1}$. Clearly a similar failure of the uncoupled harmonic oscillator approximation must also apply to the OD$^{1-4}_{\mathrm{mol-surf}}$ mode. Here the lowest transition (first to second excited state) has a frequency of $\approx$ 153\,cm$^{-1}$. Two final features are striking about the calculated wavefunctions and eigenvalues: (a) the ground, first and second excited states are sufficiently close in energy that, particularly at the elevated annealing temperatures shown in Figure \ref{SFG_data2}b, both excited states are likely populated and (b) both excited states are delocalized over the entire width of the potential. If this 1D model of the OD$^{1-4}_{\mathrm{mol-surf}}$ and OD$^{1-4}_{\mathrm{surf}}$ vibrations captures the surface physical chemistry, this result suggests that during sample thermal treatment a significant portion of interfacial deuterons are spatially delocalized (although this effect is damped at the 135 K temperature of the measurement). Because this delocalization is a result of the anharmonicity of the OD$^{1-4}_{\text{surf}}$ potential, and because it would be expected to be larger for protons than deuterons, we expect that for measurements with \ce{H2O} the degree of delocalization would only increase. We are currently working on addressing both the anharmonicity of these vibrations and their possible delocalization at moderate temperatures experimentally. 

We finally note in passing that the exceptionally low dissociation barrier of \ce{D2O} on the (1\=102) surface relative to the (0001), in addition to having interesting consequences for surface vibrational spectroscopy, is consistent with our qualitative experimental result that substantially lower dosing times are required to prepare the former surface than the latter.


\section{Summary and Conclusions\label{sum}}

We have described water adsorption and dissociation on the oxygen terminated $\alpha$-Al$_{2}$O$_{3}$(1\=102) surface using both experiment, \textit{i.e.}\ TPD and VSF spectroscopy, and theory, \textit{i.e.}\ periodic-slab DFT calculations. Combining these experimental approaches allows us to produce an oxygen terminated (1\=102) surface, adsorb a mixed monolayer of \ce{D2O} and demonstrate that with increasing temperature molecularly adsorbed \ce{D2O} appears to desorb while dissociatively, kinetically stabilized, adsorbed \ce{D2O} remains. 

Our DFT based surface model was set up in order to derive a microscopic picture of low water coverage $\alpha$-\ce{Al2O3}(1\=102) surface chemistry and allows a detailed, microscopic view of this surface speciation. We show that, in the limit of low coverage, \ce{D2O} adsorption happens via the ``1-4'' dissociation channel and that this configuration in energetically stabilized by a strong hydrogen bond donated by one OD group to a surface oxygen (while the oxygen in molecular \ce{D2O} interacts with a surface Al). This stabilization is sufficiently strong that, while the diffusion of molecular \ce{D2O} or the dissociated deuteron can occur under the conditions of our experiment thermodynamics dictates that other possible configurations should be extremely rare.

The DFT based model makes it possible to explain why we observe only two, of a possible four, resonances of molecular/dissociated \ce{D2O} in the 1-4 configuration. Evidently the dissociation barrier, \textit{i.e.}\ the barrier separating the OD$^{\text{1-4}}_{\text{mol-surf}}$ and OD$^{\text{1-4}}_{\text{surf}}$ species, is sufficiently small, and the OD transition dipole sufficiently parallel to the dissociation coordinate, that both vibrations are strongly anharmonic in character and shifted to frequencies well below our spectral window. To estimate the size of this effect we have solved the vibrational Schr\"odinger equation along the dissociation coordinate. In addition to quantifying that the fundamental OD stretch transitions of the OD$^{\text{1-4}}_{\text{mol-surf}}$ and OD$^{\text{1-4}}_{\text{surf}}$ species should be red shifted by $> 1500 \text{cm}^{-1}$ from their frequencies if they were uncoupled harmonic oscillators, this result suggests the deuterons that participate in both species may, at moderate temperatures, be delocalized. Such delocalization, if confirmed by higher level theoretical treatments and experiment, has significant implications for understanding the reactivity of the $\alpha$-\ce{Al2O3}(1\=102) to water and aqueous phase solutes. 

\section*{Acknowledgements}
\noindent We thank the Deutsche Forschungsgemeinschaft for support of this work through Collaborative Research Center 1109 \textit{Understanding of Metal Oxide/Water Systems at the Molecular Scale: Structural Evolution, Interfaces and Dissolution.}

\section*{Electronic Supplementary Information}
\noindent details of sample preparation and characterization, details of VSF spectrometer, review of detailed theory connection measured vibrational sum frequency intensity and molecular orientation and additional parameters describing computed molecular structure.

\clearpage

\bibliographystyle{ieeetr}

\clearpage

\beginsupplement

\section{\textit{Electronic Supplementary Information:} Characterization of Water Dissociation on $\alpha$-Al$_{2}$O$_{3}$(1\=102): Theory and Experiment}

\subsection{Experimental Details}
\subsubsection{Sample Preparation and Characterization}\label{s:samp}
To mount samples we followed previous workers and created a \textit{sandwich} of two $\alpha$-Al$_{2}$O$_{3}$(1\=102) crystals around a piece of 0.01 mm thick Tantalum foil secured by Tantalum clips \cite{elam98}. Mounting in this manner allows straightforward control of sample temperature between 130 and 1200 K by a combination of liquid N$_{2}$ cooling and resistance heating. The $\alpha$-Al$_{2}$O$_{3}$(1\=102) crystals we used in this study were 10 x 15 x 0.5 mm$^3$ and polished on one side to a roughness $< 0.5\ \mbox{nm}$ (as purchased from Princeton Scientific Corp).

Before mounting the crystal in the UHV chamber we placed it for 30 minutes in an ultrasonic bath with methanol, dried it with N$_{2}$ and rinsed it with milli-pure water for 30 minutes. Mounting the crystal after this procedure in the UHV chamber with no further preparation produces a surface that still shows carbon contamination in Auger Electron spectroscopy (AES). To remove the carbon, we sputtered the sample with 1.5 KV and 3 x 10$^{-5}$mbar Argon for 30 minutes at multiple spots.  As sputtering at these voltages has been shown to produce oxygen vacancies on alumina surfaces \cite{niu00}, we next annealed the sample at 1040 K for 30 minutes in an atmosphere of 5 x 10$^{-6}$ mbar of Oxygen. As shown in Figure \ref{pic2}, this treatment leads to a sample that, when analyzed with low energy electron diffraction (LEED), produces a sharp (1$\times$1)  diffraction pattern and shows no carbon contamination in AES measurements. 
\begin{figure}[H]
\begin{minipage}{0.3\textwidth}
\centering
\includegraphics[width=\textwidth]{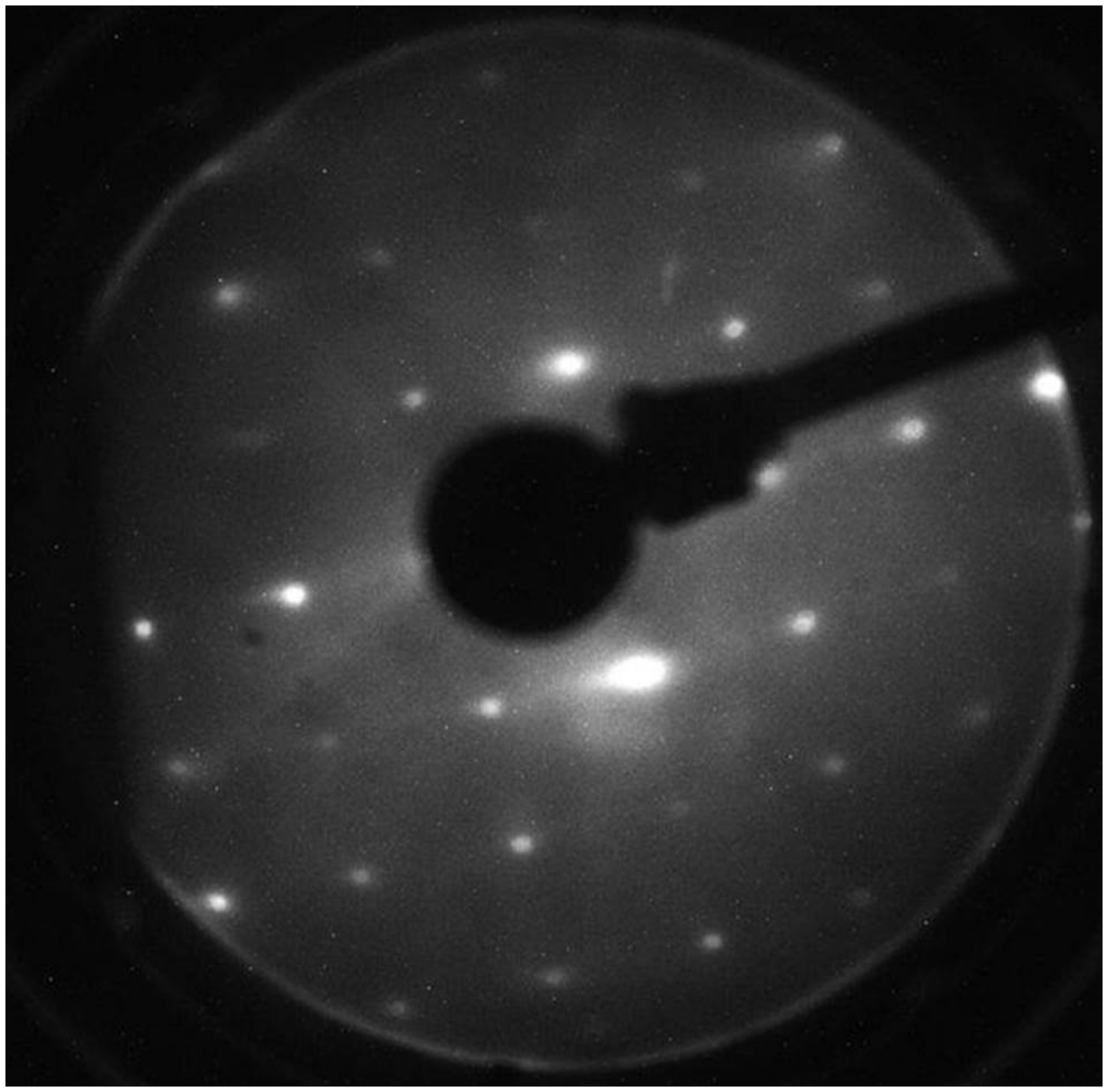}
\end{minipage}\hfill
\begin{minipage}{0.7\textwidth}
\centering
\includegraphics[angle=-90,width=1.1\textwidth]{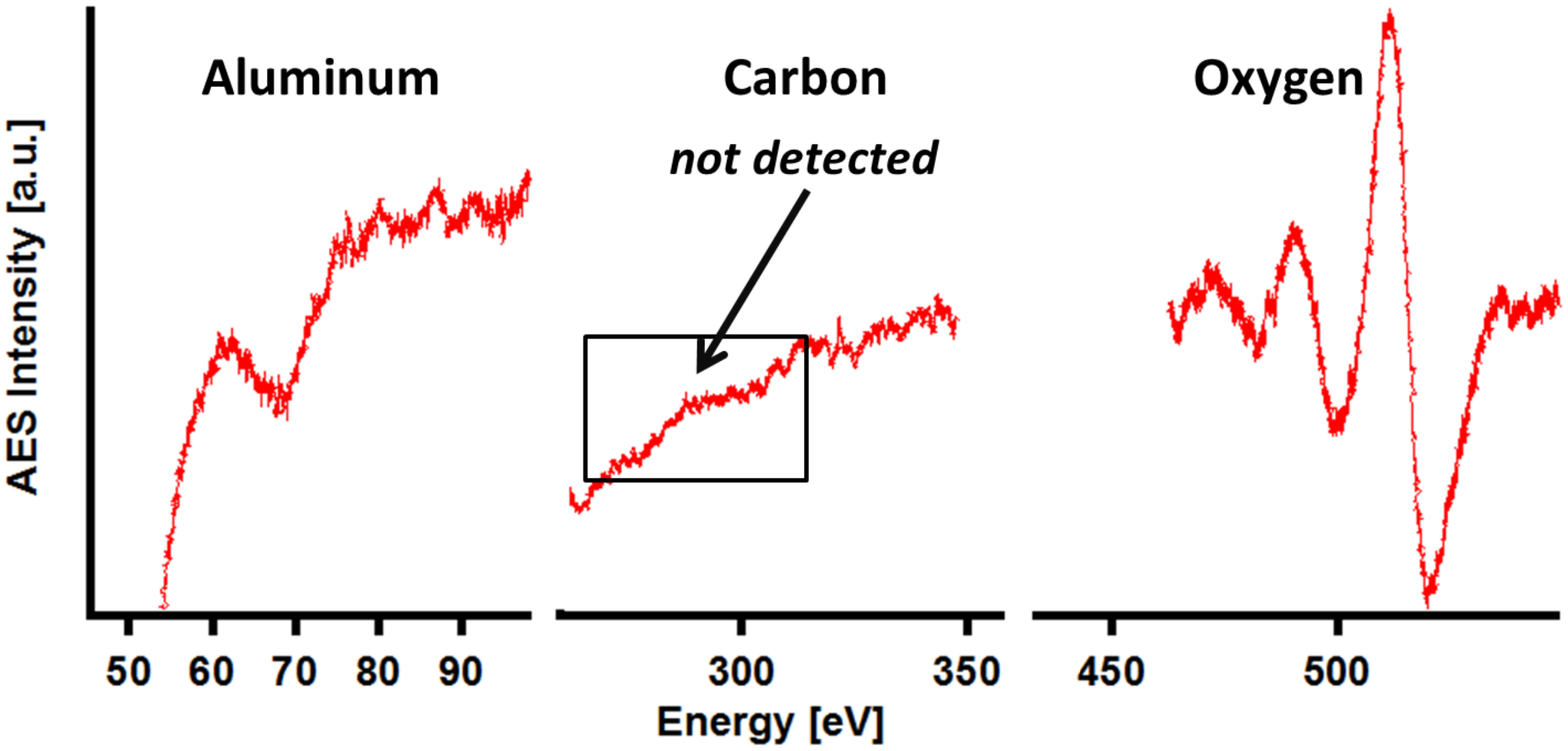}
\end{minipage}
\caption[pic2]{left: $(1\times1)$-LEED pattern taken at 120 eV and right: Auger spectra of a sputtered and annealed $\alpha$-Al$_{2}$O$_{3}$(1\=102) crystal.}
\label{pic2}
\end{figure}

This sample preparation procedure was adopted from the previous work of Trainor et. al\cite{Trainor02}, who also generated a surface that gave a well defined $(1\times1)$ LEED pattern. Other prior studies have shown that sputtering and annealing at high temperatures in the \emph{absence} of oxygen produces a surface with a (2$\times$1) LEED pattern \cite{schildbach1993} that is stable during the subsequent adsorption of water. Given that high energy Ar-ion sputtering is known to create oxygen vacancies on $\alpha$-Al$_{\text{2}}$O$_{\text{3}}$ surfaces \cite{niu00}, we conclude that a high temperature treatment in the absence of oxygen creates a nonstoichiometric, oxygen deficient, surface termination that is  is not the thermodynamically stable surface at moderate temperatures in UHV.  

\subsubsection{Temperature Measurement and Control} 
To measure sample temperature (T$_{\text{Al}_{\text{2}}\text{O}_\text{3}}$), Chromel/Alumel thermocouples were attached to the crystal edges with ceramic glue (Cerabond 605) and connected to a temperature controller (Model 340, Lakeshore) which was also connected to a resistance heating unit. We verified that the sample was heated homogeneously (temperature differences in the plane of the sample surface were $< 5$ K) by attaching thermocouples on opposite sides of the sample.

\subsubsection{Water dosing on $\alpha$-Al$_{2}$O$_{3}$(1\=102)}

To dose the clean $\alpha$-\ce{Al2O3}(1$\bar{\text{1}}$02) surface D$_2$O was seeded in helium by bubbling He through a deuterated water reservoir. To minimize adsorption of D$_2$O on the tubes of the gas system before reaching the nozzle in our molecular beam source, all tubing was heated to $120^{\circ}$C. All measurements described below were performed on samples prepared using a nozzle temperature (i.e.\ $\text{T}_{\text{Nozzle}}$) of 860 K, which corresponds to a D$_{2}$O kinetic energy of $\approx 0.6$ eV per molecule.  For dosing water on the sample we used following procedure:
	\begin{itemize}
		\item[(i)]We start dosing with the MBS at T$_{\text{Al}_{\text{2}}\text{O}_{\text{3}}}= 400$ K, cooling with a 10 K/min ramp to 130 K while dosing.
		\item[(ii)]Align the VSF spectrometer on the large free OD signal from ice by overlapping the visible and infrared beams spatially and temporally (a precondition for the VSF measurement). 
		\item[(iii)]Anneal the surface with a ramp of 100 K/min to 185 K or higher to remove the great majority of all molecularly adsorbed water and return the sample to 135 K for VSF characterization.
	\end{itemize}

\subsubsection{Technical Details of VSF Spectrometer}
In our VSF spectrometer we employ a Legend Elite Duo amplifier system (Coherent) with a pulse energy of 8 mJ, a pulse length of 50 fs, a spectrum centered at 800 nm and a repetition rate of 1 KHz. 5 of the 8 mJ output of the amplifier is sent to a TOPAS (Coherent) to generate broadband IR laser pulses. Those pulses (bandwidth 190 - 230 cm$^{\text{-1}}$ (FWHM)) have energies of 30 - 40 $\mu$J. The VIS laser in the VSF interaction is generated by narrowing the remaining 800 nm pulse after the TOPAS  in a homebuilt pulse shaper to obtain pulses of 25 $\mu$J and a FWHM of $\approx$ 8 cm$^{\text{-1}}$. The incident angles in our setup were 75/70 degrees, with respect to the surface normal, for VIS/IR. Our experimental configuration was such that both incident beams and the emitted VSF were co-planar and this plane was perpendicular to the plane of the surface. All spectrum shown were collected using the polarization combination of \textit{ppp} (VSF/VIS/IR) in which \textit{p} indicates parallel to the plane of incidence. Polarization control of all incident and detected fields was accomplished by using $\lambda$/2 wave plate, polarizer, $\lambda$/2 wave plate combinations. The emitted sum frequency field was dispersed using a grating (1800 g/mm) and detected with an ICCD camera from Princeton Instruments. 

\subsubsection{VSF Data Analysis}
To analyze the VSF data we first divided the collected signal by its corresponding nonresonant signal (to account for the frequency dependence of the incident infrared beam) and then fit the data using equations \ref{e:line1} and \ref{e:line2} and the Levenberg-Marquardt algorithm as implemented in the analysis program Igor Pro (Wavemetrics). Following prior workers we constrained this fit by taking the phase of the nonresonant contribution to be zero.  We further assumed that the phase of each resonance is temperature independent.  Treating the line width ($\Gamma$), center frequency ($\tilde{\nu}_q$) and amplitude ($|A_{q}|$) of each resonance as fit parameters leads to a description of the data in which both center frequency and line width are relatively constant as a function of temperature while amplitude varies greatly.  Attempts to fit the data with initial guesses for amplitude that vary over several orders of magnitude and phases between 0 and 2$\pi$ suggests that the resulting fit is relatively insensitive to these choices.  Initial guesses for $\tilde{\nu}_{q}$ that vary by $> 35\mbox{ cm}^{-1}$ from the tabulated results lead to converged non-physical solutions.
\begin{equation}\label{e:line1}
I_{\text{VSF}} \propto \left| \chi_{\text{eff}}^{(2)} \right|^2 \end{equation}
\begin{equation}\label{e:line2}
\chi_{\text{eff}}^{(2)} \propto \left|A_{\text{NR}} \right|\ e^{\text{i}\phi_{\text{NR}}} +\ \sum_{q} \frac{|A_{q}|e^{\text{i}\phi_{\text{R}}}}{\tilde{\nu}_{\text{IR}}-\tilde{\nu}_q + i\Gamma_q} 
\end{equation}

\begin{table}
	\begin{center}
		\begin{tabular}{ccccc}
			\textsf{OD species} 		& $\tilde{\nu}_q$ \textsf{(cm}$^{-1}$\textsf{)}	& $\Gamma$ \textsf{(cm}$^{-1}$\textsf{)}	&	$|A_q|$\textsf{(a.u.)}	&	\textsf{$\phi_{\text{R}}$(}$\pi$\textsf{)}  \\
			\midrule 
			 OD$_{\mathrm{ice, 135K}}$        & 2723 & 12	&	400	& 0.5 \\
			 OD$_{\mathrm{mol, ads,185K}}$        & 2732 & 26	&	33	& 0.1 \\
			 OD$_{\mathrm{diss, ads,185K}}$        & 2772 & 37 	&	28	& 0.2 \\ 
			 
		\end{tabular}
			\vspace{0.2cm}
		\caption{Results from applying the line shape model in the using the procedure described above for an ice sample and a sample at 185 K.}
		\label{para}
	\end{center}
\end{table}

\subsection{Linking I$_{\text{VSF}}$ and OD angle}
Monitoring the intensity of the emitted VSF signal as a function of incident IR frequency allows us to recover a vibrational spectrum of all OD groups at the surface. This observed VSF spectral response is a function of the microscopic nonlinear susceptibility; the incident beam angles; refractive indices of both bulk phases and the interface; and molecular orientation. The manner in which these properties of the interfacial system relate has been well described in previous studies \cite{guyo86, zhu87, hein91, bain91, wang05}, here we only review the aspects of this theory relevant to understanding the VSF data reported here. 

Assuming the emitted sum frequency field originates from a polarization sheet at the $\alpha$-Al$_{2}$O$_{3}$/vacuum interface and solving Maxwell's equations gives, 

\begin{eqnarray}\label{e:sfg_sig}
I_{\text{VSF}}(\tilde{\nu}_{\text{IR}}) = \frac{8 \pi^{3} \tilde{\nu}^2_{\text{VSF}}}{c^3 cos^{2}\beta_{\text{VSF}}} \left| \chi_{\text{eff}}^{(2)} \right|^2 I_{\text{VIS}}I_{\text{IR}}(\tilde{\nu}_{\text{IR}})
\end{eqnarray}
in which $I_{\text{VSF}}(\tilde{\nu}_{\text{IR}})$ is the intensity of the emitted sum frequency field (and is a function of the frequency of the incident infrared), $\tilde{\nu}_{i}$ the frequency of the $i^{th}$ field, c the speed of light in vacuum, $\beta_{\text{VSF}}$ the angle of the wave vector of the reflected SF field with respect to the surface normal, $I_{\text{VIS}}$ the intensity of the visible field and $I_{\text{IR}}(\tilde{\nu}_{\text{IR}})$ is the intensity of the incident infrared, which is frequency dependent.  All of these parameters, except the $\chi^{(2)}_{\text{eff}}$, are either under the control of the experimentalist or are physical constants. 

$\chi^{(2)}_{\text{eff}}$, the macroscopic nonlinear susceptibility, contains all information about the nonlinear optical properties of the sample. Because our surface has macroscopic $C_{\infty v}$ symmetry, 7 of the 27 elements of $\chi^{(2)}_{\text{eff}}$ are nonzero and only 4 are independent. In a laboratory reference frame where (x,y) is the plane of the surface and z the surface normal, these are  $\chi_{zzz}^{(2)}$,  $\chi_{xzx}^{(2)} = \chi_{yzy}^{(2)}$, $\chi_{xxz}^{(2)} = \chi_{yyz}^{(2)}$ and  $\chi_{zxx}^{(2)} = \chi_{zyy}^{(2)}$. If we assume that all beams propagate in the x-z plane and $s$ indicates polarization perpendicular and $p$ parallel to the x-z plane, we can write explicit expressions relating the experimentally controllable parameters, beam polarizations and angles, to the $\chi^{(2)}_{\text{eff}}$. For $\chi^{(2)}_{\text{eff,ssp}}$ and $\chi^{(2)}_{\text{eff,ppp}}$ the relevant equations are, 
\begin{eqnarray}
\chi_{\text{eff,ssp}}^{(2)} & = & L_{yy}(\tilde{\nu}_{\text{VSF}}) L_{yy}(\tilde{\nu}_{\text{VIS}}) L_{zz}(\tilde{\nu}_{\text{IR}})\sin\beta_{\text{IR}} \chi_{yyz}^{(2)} \\
\chi_{\text{eff,ppp}}^{(2)} & = & -L_{xx}(\tilde{\nu}_{\text{VSF}})L_{xx}(\tilde{\nu}_{\text{VIS}})L_{zz}(\tilde{\nu}_{\text{IR}})\cos\beta_{\text{VSF}}\cos\beta_{\text{VIS}}\sin\beta_{\text{IR}}\chi^{(2)}_{xxz} \nonumber\\
					&  & -L_{xx}(\tilde{\nu}_{\text{VSF}})L_{zz}(\tilde{\nu}_{\text{VIS}})L_{xx}(\tilde{\nu}_{\text{IR}})\cos\beta_{\text{VSF}}\sin\beta_{\text{VIS}}\cos\beta_{\text{IR}}\chi^{(2)}_{xzx} \nonumber\\
					&  & +L_{zz}(\tilde{\nu}_{\text{VSF}})L_{xx}(\tilde{\nu}_{\text{VIS}})L_{xx}(\tilde{\nu}_{\text{IR}})\sin\beta_{\text{VSF}}\cos\beta_{\text{VIS}}\cos\beta_{\text{IR}}\chi^{(2)}_{zxx} \nonumber\\
					&  & +L_{zz}(\tilde{\nu}_{\text{VSF}})L_{zz}(\tilde{\nu}_{\text{VIS}})L_{zz}(\tilde{\nu}_{\text{IR}})\sin\beta_{\text{VSF}}\sin\beta_{\text{VIS}}\sin\beta_{\text{IR}}\chi^{(2)}_{zzz}\nonumber
\end{eqnarray}
in which $\sin\beta_{k}$ is the incident angle for the beam k and $L_{ij}(\tilde{\nu})$ are Fresnel coefficients. These coefficients can be expressed as,
\begin{eqnarray}
	L_{xx}(\tilde{\nu}) & = & \frac{2\cos\gamma_{\tilde{\nu}}}{\cos\gamma_{\tilde{\nu}} + n_{Al_{2}O_{3}}(\tilde{\nu})\cos\beta_{\tilde{\nu}} } \\
	L_{yy}(\tilde{\nu}) & = & \frac{2\cos\beta_{\tilde{\nu}}}{\cos\beta_{\tilde{\nu}}+n_{Al_{2}O_{3}}(\tilde{\nu})\cos\gamma_{\tilde{\nu}}} \\
	L_{zz}(\tilde{\nu}) & = & \frac{2n_{Al_{2}O_{3}}(\tilde{\nu})\cos\beta_{\tilde{\nu}}}{\cos\gamma_{\tilde{\nu}} + n_{Al_{2}O_{3}}(\tilde{\nu})\cos\beta_{\tilde{\nu}}} \left(\frac{1}{n^{\prime}(\tilde{\nu})}  \right)
\end{eqnarray} 
in which $\beta_{\tilde{\nu}}$ is the incident angle of the beam at frequency $\tilde{\nu}$, $n_{Al_{2}O_{3}}(\tilde{\nu})$ is the refractive index of $\alpha$-Al$_{2}$O$_{3}$ at frequency $\tilde{\nu}$, $\gamma_{\tilde{\nu}}$ is the refracted angle of the field at frequency $\tilde{\nu}$ (i.e. $\sin\beta_{\tilde{\nu}} = n_{Al_{2}O_{3}}(\tilde{\nu})\sin\gamma_{}$) and $n^{\prime}$ is the interfacial refractive index. Note that these coefficients depend only on the experimental geometry (\textit{e.g.}\ the angle of incoming IR and VIS beams) and the refractive index of both bulk phases and the interface but have no contribution from material nonlinear optical properties. 

The macroscopic nonlinear susceptibility in the laboratory frame $ \chi_{ijk}^{(2)}$ can be connected to the molecular hyperpolarizability $\alpha_{i^{\prime}j^{\prime}k^{\prime}}$,
\begin{equation}\label{e:trans}
\chi_{ijk}^{(2)} = \frac{1}{2\epsilon_{0}} N_{s}\langle R_{ii^{\prime}}R_{jj^{\prime}}R_{kk^{\prime}}\rangle\alpha_{i^{\prime}j^{\prime}k^{\prime}}
\end{equation}
in which $\langle R_{ii^{\prime}}R_{jj^{\prime}}R_{kk^{\prime}}\rangle$ is the ensemble averaged transformation matrix between molecular and laboratory coordinates in the \textit{slow motion limit}: the limiting case in which the angle of individual OD groups with respect to the surface normal ($\theta$) does not change on the timescale of the inverse line width of the measured spectral response. It is unlikely that our system rigorously meets this condition but, as has been discussed in the previous literature in detail (see \cite{wang05} and refs therein), for physically realistic motion of OD groups the effect on relative OD intensities with changing experimental geometry is likely to be small (note here that all measurements are performed at 135 K). Given this assumption, evaluating the $\langle R_{ii^{\prime}}R_{jj^{\prime}}R_{kk^{\prime}}\rangle$ matrix for each of the relevant $\chi^{(2)}_{ijk}$ terms gives,  

\begin{eqnarray}
\chi_{yyz}^{(2)} & = & \chi_{xxz}^{(2)} = \frac{1}{2 \epsilon_0} N_s \alpha_{z^{\prime}z^{\prime}z^{\prime}}^{(2)} \left[ \left\langle sin^2 \theta \ cos\theta    \right\rangle (1-r)+2r \left\langle cos\theta   \right\rangle \right]\nonumber\\
\chi_{xzx}^{(2)} & = & \chi_{zxx}^{(2)} = \frac{1}{2\epsilon_0} N_s \alpha_{z^{\prime}z^{\prime}z^{\prime}} \left(\langle\cos\theta\rangle - \langle\cos^{3}\theta\rangle\right)(1-r)    \nonumber\\
\chi_{zzz}^{(2)} & = & \frac{1}{\epsilon_0}N_s \alpha_{z^{\prime}z^{\prime}z^{\prime}}\left[r\langle\cos\theta\rangle + \langle\cos^{3}\theta\rangle \left(1-r\right)   \right]    \label{e:orient}
\end{eqnarray}
in which $N_{s}$ is the number of molecules at the surface, $\alpha_{z^{\prime}z^{\prime}z^{\prime}}^{(2)}$ is  the hyperpolarizability along the molecular axis $z^{\prime}$, $r$ is is the hyperpolarizability ratio ($r= \frac{\alpha_{y^{\prime}y^{\prime}z^{\prime}}}{\alpha_{z^{\prime}z^{\prime}z^{\prime}}}$) and $\theta$ is the angle with respect to the surface normal. To calculate the relationship between $I_{\text{VSF}}$ and $\theta$ as presented in the text and below we have assumed that all non-hydrogen bonded OD have the same $\alpha_{z^{\prime}z^{\prime}z^{\prime}}$ and $r$ and used the values for these quantitates from our prior publication \cite{tong13}.  Because we desire insight only into the \emph{relative} change in I$_{\text{VSF}}$ as a function of OD orientation, knowledge of the surface density of OD groups is not required.

\subsection{Computational Details}
\subsubsection{Additional Structural Parameters of Adsorbed D$_2$O}
As discussed in the text we principally consider eight possible OD groups in this paper. Two of these are contained in an intact D$_{2}$O molecule adsorbed in the 1-2 configuration, two in an intact D$_{2}$O molecule adsorbed in the 1-4 configuration, two in a dissociatively adsorbed D$_{2}$O molecule in the 1-2 configuration and two in a dissociatively adsorbed D$_{2}$O molecule in the 1-4 configuration. In the manuscript the angle formed by each of these eight OD groups with the surface normal is tabulated. We here provide additional structural parameters.
\begin{table}[htp]
\caption{Structural parameters from the four structures illustrated in Figures 4 and 5 in the manuscript. Al-O$_{\text{D}_{2}\text{O}}$ indicates the bond between a surface Al and the oxygen in an adsorbed D$_{2}$O. Al-O$_{\text{D}_{2}\text{O}}$ angle is the angle of the Al-O$_{\text{D}_{2}\text{O}}$ bond vector with respect to the surface normal.}
\begin{center}
\begin{tabular}{lcccc}
\textsf{Structures} & \textsf{1-4 molec} & \textsf{1-4 dissoc} & \textsf{1-2 molec} & \textsf{1-2 dissoc} \\
\midrule
Al-O$_{\text{D}_{2}\text{O}}$ bond length [\AA] & 1.94 & 1.80 & 2.00 & 1.75 \\
Al-O$_{\text{D}_{2}\text{O}}$ angle [$^{\circ}$] & 35.5 & 37.1 & 13.1 & 22.2 \\
OD$_{\text{ads}}$ bond length [\AA] & 0.97 & 0.97 & 0.98 & 0.97 \\
OD$_{\text{surf}}$ bond length [\AA] & 1.54 & 1.03  & 2.03 & 0.98 

\end{tabular}
\end{center}
\label{default}
\end{table}%

\clearpage

\bibliographystyle{ieeetr}

\end{document}